\documentclass[twoside,11pt]{article}

\usepackage{jmlr2e}
\usepackage{xcolor}
\usepackage{amsmath}
\usepackage{subfig}
\usepackage[ruled,vlined]{algorithm2e}
\usepackage{array}
\usepackage[acronym]{glossaries}
\usepackage{multicol}
\usepackage{tikz}

\newcommand*\circled[1]{\tikz[baseline=(char.base)]{% <---- BEWARE
            \node[shape=circle,draw,inner sep=0.2pt] (char) {#1};}}

\newtheorem{defn}{Definition}
\ShortHeadings{EEG Signal Processing for Seizure Detection}{Grant and Islam}
\firstpageno{1}

\begin{document}

\title{ EEG Signal Processing using Wavelets for Accurate Seizure Detection through Cost Sensitive Data Mining}

\author{\name Paul G. Grant         \email pgrant@csu.edu.au \\
   %      and \\
       % \AND
        \name Md Zahidul Islam       \email   zislam@csu.edu.au          \\
        \addr School of Computing and Mathematics\\
       Charles Sturt University\\
       Bathurst, NSW 2795, Australia}
   %    \AND
   %    \name Michael I.\ Jordan \email jordan@cs.berkeley.edu \\
   %    \addr Division of Computer Science and Department of Statistics\\
    %   University of California\\
   %    Berkeley, CA 94720-1776, USA}

\editor{TBA}

\maketitle

\begin{abstract}%   <- trailing '%' for backward compatibility of .sty file
Epilepsy is one of the most common and  yet diverse set of chronic neurological disorders. This excessive or synchronous neuronal activity is termed  ``seizure". Electroencephalogram  signal processing plays a significant role in detection and prediction of epileptic seizures. In this paper  we introduce an approach that relies upon the properties of wavelets for seizure detection. We utilise 
 the Maximum Overlap Discrete Wavelet Transform which enables us to  reduce signal \textit{noise}. Then from the variance exhibited in wavelet coefficients we develop connectivity and  communication efficiency between the electrodes as these properties differ significantly during a seizure period in comparison to a non-seizure period. We use basic statistical parameters derived from the reconstructed noise reduced signal, electrode connectivity and the  efficiency of information transfer to build the attribute space. %We have relied heavily upon open source software in regards to the wavelet transforms and the application of the classifier to the transformed data.%
 We have utilised data that are publicly available to test our method that is found to be significantly better than some existing approaches.
 
 %\textit{Conditional upon Journal acceptance, the associated R code will be made available via Github}

\end{abstract}

\begin{keywords}
 EEG, Graph Theory, MODWT, Wavelet variance, Decision trees 
\end{keywords}

\section{Introduction}\label{sec:intro}
Epilepsy is described by recurring seizures caused by
abnormal activity in the brain. This maybe considered  a chronic disorder of the central nervous system that exposes individuals to experiencing recurrent seizures.  A seizure is a transient irregularity in the brain's electrical processes that may also produce disruptive physical symptoms \citep{shan:2010}.
Electroencephalogram (EEG) is a signal which represents
the electrical activity of  neurons within the brain.
The signal is acquired from the surface of the scalp and these signals are non-stationary,  that is its statistical properties  vary over time.
 The important frequencies from
the physiological viewpoint lie in the range of 0.1 to 30
Hz. \citep{shres:2019}. A common classification problem is seizure detection, where non-seizure and seizure EEG records of patients need to be identified. \par
\begin{figure}[ht]
    \centering
    \includegraphics[scale=.3]{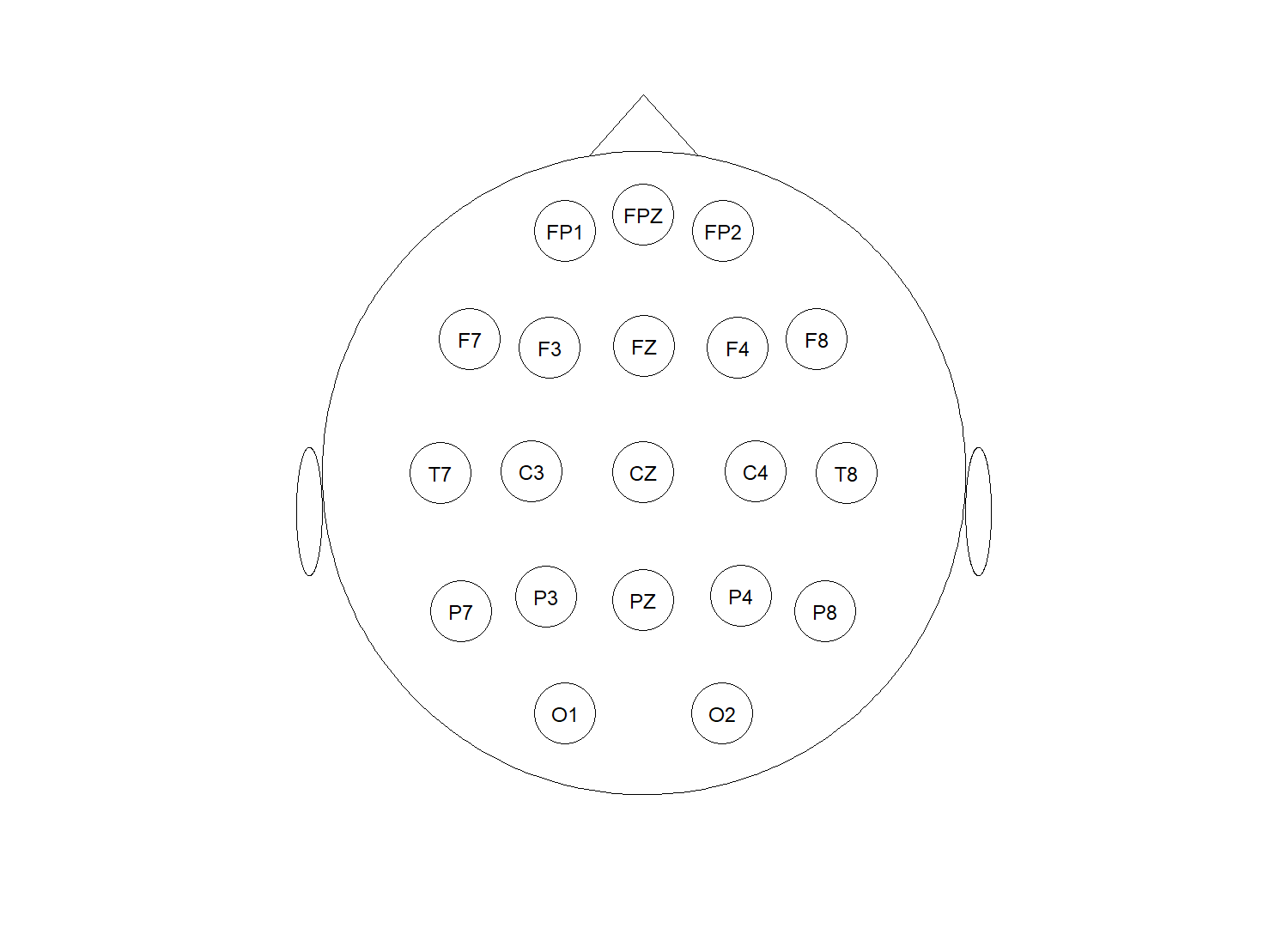}
    \caption{ 10 - 20 Electrode positioning; \textit{developed from \cite{Helw:2018}.}  }
    \label{fig:EEGc}
\end{figure}

 It is not an onerous  task to find a number of papers and differing methods that seek to determine  the onset of seizure.
The nearest-neighbour classifier was used on EEG features extracted in both time and frequency domains to detect the onset of epileptic seizures \citep{qugot:97}.  Also a method using various derived statistical parameters as extracted features has been proposed \citep{sidd:2018}. One problem with these methods is that all the information in the signal is being considered. Some of that information may simply be noise and may also contain specific frequency bands  that  are not related to the event that is to be classified.
 In another study the application of a 
method based on energy extraction at various  frequency sub-bands using the wavelet transform was implemented \citep{Jacob:2018}. This avoids processing redundant data by selecting the required sub-bands / wavelet decomposition levels which contain relevant information. Statistical parameters of the energy at each level were derived and used as attributes for a Support Vector Machine (SVM) classifier. Using this method with statistical parameters of energy, at different frequency levels, and in this case with 21 electrodes (arranged in the international ``10-20" format, as an example see Figure \ref{fig:EEGc}.), the amount of computation required  increases quickly if we require additional  frequency bands to better segregate the signal or additional statistical parameters to fully describe the energy feature. This results from the need to compute the statistical parameters for the energy within each additional sub-band/wavelet decomposition level required and of course for the signal from each electrode.   \par
The EEG signal contains several spectral components.  Wavelet analysis allows  frequency  separation of such components \citep{percival_walden_2000}. Use of Wavelets in EEG analysis is quite often seen and is well described, with the Discrete Wavelet transform (DWT)  frequently encountered \citep{Faust:2015}. (For brief overview of DWT see Appendix \ref{App:DWT}).  The wavelet transform is 
effective for representing different aspects of non-stationary signals such as trends, discontinuities, and where signal patterns are recurring,  here other signal processing approaches 
are unlikely to be as effective \citep{Daub:1990}. Some limitations of the DWT are the requirement of the signal to be of length $2^J$ where $J \in \mathbb{Z}^+$ and is the upper level of wavelet decomposition. Also as $J$ increases the  number of wavelet coefficients within each additional level systematically reduces, so to derive meaningful statistical parameters at these upper levels of decomposition might not always be possible. \par %One advantage of the wavelet transform is that it has a varying
%window size, being wide at low frequencies and then narrowing at high frequencies. This permits  wavelet analysis of  a signal at different frequencies with different resolutions. Wavelets provide good frequency resolution and relatively poor temporal resolution at low frequencies and conversely good temporal resolution and relatively poor frequency resolution at high frequencies. 
%Use of Wavelets in EEG analysis is quite often seen and is well described, with the Discrete Wavelet transform (DWT, see Appendix \ref{App:DWT}) frequently encountered \citep{Faust:2015}.\par
Our proposed method minimises the use of non-required frequency ranges within the signal, reconstructs the signal to its original length and overcomes some of the possible shortfalls of the DWT.  It also uses the interaction between subsets of electrodes during an epoch. This interaction or connectivity can be seen to vary significantly between the non-seizure and seizure periods.  We utilise statistical parameters from  our reconstructed signal and electrode connectivity as  extracted features for classification. 
The organisation of this paper is as follows; Section 2 outlines some related works to our proposed method, Section 3 presents our method, Section 4 outlines experimental results together with compared works. Concluding remarks are presented in Section 5, followed by an Appendix which includes a Glossary.

%{\noindent \em Remainder omitted in this sample. See http://www.jmlr.org/papers/ for full paper.}
\section{Related Work} \label{sec:RelW}

Time and frequency domains are commonly used in feature extraction methods for EEG signals. To analyse the frequency domain features, the Discrete Fourier Transform (DFT) has been applied, followed by a form of power spectral density analysis of the EEG signals \citep{Lee:2014}. Decomposing a signal in terms of its frequency content, using the DFT which relies on  sinusoids, results in fine resolution within the frequency domain. However the Fourier series representation is not very effective at  time resolution \citep{Vikash:2015}.  Another method of decomposing signals is the DWT, where in  a  wavelet representation, we represent the signal in terms of functions that are localised both in time and frequency. Here we have an adaptive time frequency window, which provides good time resolution at high frequencies, and good frequency resolution at low frequencies \citep{Polikar:1994}.

 \cite{sidd:2018} generated nine statistical parameters from EEG data to  extract time domain  features  and reduce dimension. The duration of data \textit{(Epoch)} was also altered,  starting at 10 sec, and reducing in steps to 0.025 secs. The classifiers chosen there were  decision forests, an ensemble of decision trees. A concern with this methodology is that one is using all the available information or frequency bandwidth within the signal. Such techniques as DFT or DWT were not used to minimise the inclusion of some sections of the signal's bandwidth, it is quite possible that additional \textit{noise}\footnote{Here we define \textit{noise} as frequency ranges unrelated to a seizure event.}  may have been introduced. Hence the  likelihood of introducing error into the classification of  the EEG signal. \par %Our proposed method  using wavelet decomposition and reconstruction, considers  relevant frequencies and minimises noise.\par

EEG signals have also been decomposed into time–frequency representations using the DWT \citep{Parv:2013, Faust:2015}. This enables the EEG signals to be decomposed into several sub-bands, this decomposition is called Multi Resolution Analysis (MRA). By Parseval's Theorem, the percentage distribution of energy features in the EEG signal may be extracted at each of these different sub-bands \citep{Omer:2013}.  The features extracted from the wavelet  coefficients at various levels or different frequency bands can be used to determine the 
characteristics of the signal. Use of the DWT continues to  exhibit  improving results on seizure detection \citep{Faust:2015}. One concern here is that with the DWT, as we increase the levels of decomposition to extract the relevant frequency bandwidths  we have  fewer wavelet detailed coefficients $\{d_{i,j}\}$,  within these higher levels of decomposition\footnote{Here $d_{i,j}$ represents the $i^{th}$ detail wavelet coefficient at the $j^{th}$ scale or level and $\mathcal{D}_j$ represents all the detail coefficients at level $j$. Similarly for the smooth wavelet coefficients, $s_{i,J}$ and $\mathcal{S}_J$.}.   That is $\{ d_{1,j}, d_{2,j}, \ldots, d_{k,j} \} \in \mathcal{D}_j $ has fewer elements as $j$ increases: $0 \le j \le J$. Hence the difficulty in deriving meaningful statistical parameters at the higher decomposition levels.  %Our proposed method which uses an inverse transform to reconstruct the original signal  after frequency selection, provides the same number of data points as in the original signal for the chosen epoch. \par

%{\color{red} UP TO HERE}

 %Wavelet entropy may be  obtained from the wavelet coefficients on sub-band  decomposition, this wavelet entropy has been used as features\citep{kuma:2010}. Similarly certain features based on DWT were obtained,  with different classifiers applied to determine epileptic EEG classification, including methods such as  artificial neural network \citep{Shanm:2019}, one nearest neighbour (1-NN) and support vector machine (SVM) \citep{Xie:2012}. \par
  
 \cite{Chen:2017} uses various different DWTs for seizure detection, also choosing from nine statistical features derived from each of the wavelet generated frequency bands then classifying (via SVM),  the multi-channel EEG recordings from \cite{PhysioNet}. The empirical evidence shown there outlines that the choice of mother wavelet does not have a significant effect  on seizure detection, however it is very sensitive to decomposition level if the features of non-seizure/seizure EEGs exhibit significant difference
in the different frequency bands. In this particular study each Epoch of EEG data used was 20 seconds in duration.  Of concern here is that many frequency bands and the resulting DWT coefficient features could be redundant, causing inaccuracy and unnecessary high computational cost. If we decompose to 6 levels: $\mathcal{D}_1, \mathcal{D}_2, \ldots, \mathcal{D}_6\; \&\; \mathcal{S}_6$ and use all the wavelet coefficients within each of these  levels to derive statistical parameters,  the result set of attributes to be used for classification becomes quite large for multivariate EEG data.\par

\begin{figure}[ht]
    \centering
    \includegraphics[scale=.35]{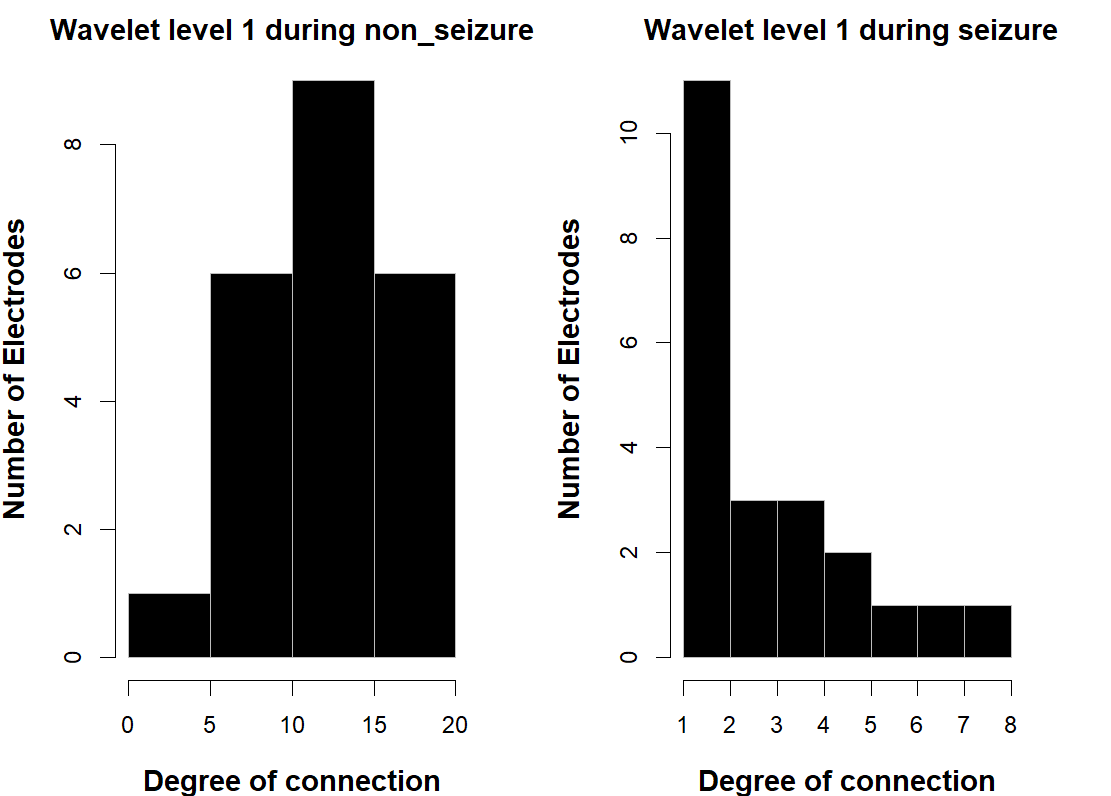}
    \caption{ Electrode Connections with correlation threshold = 0.125}
    \label{fig:WaveConn}
\end{figure}

While use of the DWT within EEG analysis is often encountered, however use of the variance within the wavelet coefficients resulting from the transform, mainly consider this wavelet variance as a selected feature only. This is usually  the standard deviation of the detail wavelet coefficients at  each decomposition scale or a linear combination of the variance  across the different wavelet decomposition scales \citep{Janj:2010, Ashok:2017}. However little if any consideration is given  to the covariance between different signals/nodes, which may be calculated similarly as the wavelet variance, this is called wavelet covariance, see definition \ref{sam:covar} Appendix \ref{Defs}, as well as Equations \ref{wavVar}, \ref{wavCoVar} and \ref{covarXY}. The covariance ( and wavelet correlation, see Equation \ref{wave:corel} ) can be used to derive electrode/node connectivity during an epoch to provide additional features for classification, especially if the connectivity between electrodes/nodes alters significantly between the non-seizure and seizure states. As an example this is  shown in Figure \ref{fig:WaveConn}, highlighting an epoch for both non-seizure and seizure states, displaying an electrode's number of connections to other electrodes. It can be seen that during the non-seizure state, there exists numerous connections between the electrodes, i.e. 6 electrodes have connections ranging from 5 to 10 other electrodes  and another 6 electrodes have connections ranging from 15 to 20 other electrodes.  Where as in comparison to non-seizure here no electrode has more than eight connections to other electrodes and only one electrode  has a maximum of eight connections at the chosen wavelet correlation threshold.
%\subsection{Issues}

The discrimination of EEG events, using statistical parameters derived from  multivariate  data  as attributes  are likely to include considerable information unrelated to the event under study, noticed as noise and  may induce error. However using the DWT in an attempt to reduce noise by selecting the required frequency bands  (wavelet decomposition levels), problems may still arise. One obvious issue is length of signal, if not a multiple of $2^J$ where $J \in \mathbb{Z}^+$, then additional components would need to be added in as extra coefficients or they may be simply omitted. Either option  may introduce bias, hence possibly confuse classification. 

\section{Our Proposed Technique}
%\input{4prob_desc3}
%\subsection{ Synopsis of  results and comparisons } \label{sub:results}
% \input{5Compar4}
As we are interested in the classification of  neurological  events,  labelled as either non-seizure or seizure states, where the data is presented in a multivariate time series format. Here we present a  novel approach for feature extraction and classification using  the inherent capabilities of the Maximum Overlap discrete Wavelet
Transform (MODWT, Appendix \ref{App:moDWT}) to:
\begin{align*}
   \blacklozenge & \quad \text{decompose and then reconstruct the EEG signal minimising the noise components} \\ %\left.\begin{array}{l}  \\ \end{array}\right.
   \blacklozenge     & 
    \left. \begin{array}{l}
    \text{construct a record of inter-electrode connectivity and} \\
    \text{develop a metric for  communication efficiency of electrodes}
    \end{array}\right\}\textit{ from wavelet covariance.}
\end{align*}

%\begin{itemize}
%    \item [$\blacklozenge$] decompose and then reconstruct the EEG signal minimising  the   \textit{noise}\footnote{Here we define \textit{noise} as frequency ranges unrelated to a seizure event.} components
%\item [$\blacklozenge$] construct a record of inter-electrode connectivity
%\item [$\blacklozenge$] and develop a measure  for  efficiency of electrode communication.
%\end{itemize}

 From these wavelet derived methods, we build our attribute space and apply an ensemble classifier to the transformed data.\par
The scenario we are considering is to have a database consisting of labelled EEG signals from a number of previous or existing patients. We subset the signal into smaller periods of time (Epochs) and label each epoch (non-seizure or seizure), determined by where the epoch occurs in time, within the signal. %Part of the signal can be labelled as seizure and the remaining part labelled as non-seizure.
We construct a model from this training data, and apply the model on incoming unlabelled EEG data, an epoch at a time, from a patient to classify/detect  non-seizure and seizure events. 

When using raw data and deriving statistical parameters as attributes, then our method if using same number of derived statistical parameters, we notice slightly more attributes per record. However, we reduce the number of attributes when compared to standard DWT decomposition methodology.
%Initially from  this training database where the seizure duration is defined for a patient, for each electrode (or signal), we subset the signal into required epoch lengths and label each epoch as either seizure or non-seizure, depending upon where the epoch exists within the signal. Our training data for each individual 

\subsection{Main Steps}\label{sub:meth}
% \subsubsection{Overview} \label{subs:Over}
We outline the main steps of our method in this subsection, then explain each of these steps in greater detail in the following subsections. With the raw data chosen then  %available within the public domain, we similarly used open source freeware. Being; R \citep{rcran}  with main packages used; WMTSA \citep{wmtsa:2017}and brainwaver \citep{brain:2012} for processing, wavelet  and statistical analysis, together with WEKA \citep{wekas} for classification. 
after subsetting the data  for each patient into the required time duration  or epoch, we begin a three step process to attain  the required attributes, then combine all into a data set followed by the 4th step, Classification.

\begin{enumerate}
\item [\textit{Step 1.}] Preparation of Attributes  from statistical parameters of reconstructed signal. %Apply  the MODWT  to decompose signal into required frequency bands. Reconstruct signal via inverse transform removing unwanted frequency bands. %From this partially reconstructed signal we extract  statistical parameter and use as attributes. %This provided us with the initial $5 \times 23 = 115$ attributes.\
\item [\textit{Step 2.}]
Preparation of Attributes from Connectivity of Electrodes %Using the MODWT  derive wavelet variance,  calculate the correlation between the electrode signals, construct an adjacency matrix, where a pair of electrodes with a correlation higher than a set threshold has a value of 1, otherwise zero. %Determine number of direct connections between  electrodes during each epoch and calculate statistical parameters, such as maximum number of connections and average number of connections, to use as attributes %This provided us with another 5 attributes. 

\item [\textit{Step 3.}] Preparation of Attributes from Global Efficiency of Electrodes %From the adjacency matrix constructed in \textit{Step 2}, calculate  global efficiency of how each electrode communicates with other electrodes  in network during the epoch.% From  this derived global efficiency, which is a  value between 0 and 1 for each node (\textit{electrode}) during an epoch, we again take statistical parameters as attributes.  
%\end{enumerate}

%\textit{From these three steps we combine the results to construct the attributes  per epoch for each individual, hence establish a new transformed data set. }

% \begin{itemize}
%     \item 10 sec epoch =  approx 360 records, with 5 $\times$ 5 $\times$ 23 = 575 attributes long + label.
%     \item 4 sec epoch  = 900 records, with 576 attributes.
%     \item 1 sec epoch  = 3600 records with 576 attributes.
% \end{itemize}
%\textit{These three steps combined together provides  a total 115 + 5 + 5 = 125  attributes.  Each step is further outlined in Algorithm \ref{Algo1} and detailed in  our experiment, section  \ref{sec:tech}. 
%Our constructed  database  of 12 patients which equates to approx 42818 records, each record of 125 attributes plus label.}

\item [\textit{Step 4.}] Assigning Class labels and Training Classifiers % From these first three steps we combine the results to construct the attributes per epoch for each individual, hence establish a new transformed data set. Then apply a cost sensitive classifier to the transformed data.% CSForest \citep{sier:2014} classifier to the transformed data. % using this data as the test set and previously constructed data from other individuals as training set.%: For each individual, remove their records  from the database to create a test set,  and use the remaining individuals as a training set to build a classification model upon. Repeat for each individual\footnote{A leave one out cross validation method}.  For classification  we used a cost-sensitive classification technique called CSForest \citep{sier:2014}, which is an ensemble of decision trees.

\end{enumerate}

 % {\color{red}Insert detailed algorithm here}
% \subsubsection{Instruction Set} \label{sub:Algo1}

\subsection{Step 1: Preparation of Attributes  from statistical parameters of reconstructed signal.} \label{sub:step1} 
\begin{figure}[ht]
    \centering
    \includegraphics[width=13.5cm, height=8.2cm]{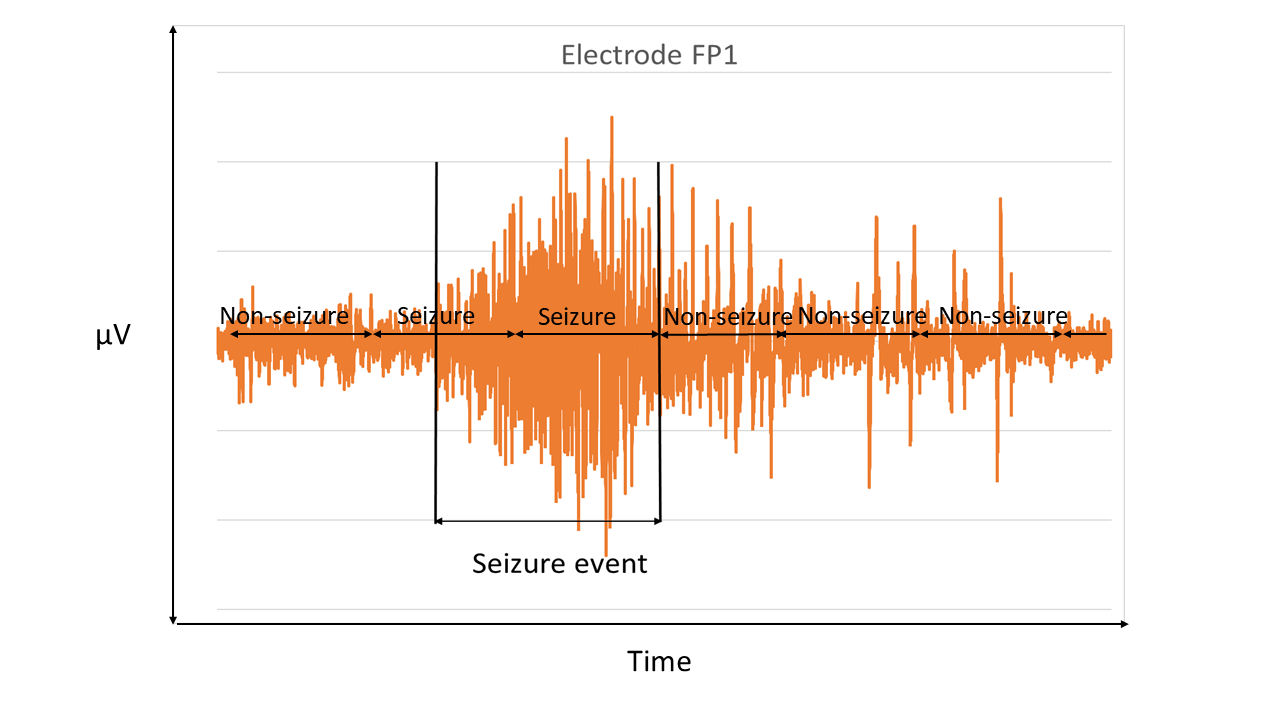}
    \caption{Data from a single electrode, with seizure duration defined and Epochs labelled.  }
    \label{fig:EEGdata}
\end{figure}

With  data, originating  from an EEG skull cap sampling method, see Figure \ref{fig:EEGc} for an example  of the standard EEG  ``10-20" skull cap electrode configuration. The data appears as a time series or signal from each electrode. The non-seizure and seizure periods are previously identified. We segment the signal into epochs of suitable duration and attach class labels. Any epoch with part of its duration within the identified seizure period is defined as seizure, elsewhere non-seizure. As an example see Figure \ref{fig:EEGdata}. Each segment along the X-Axis within a left and right arrow represents a epoch. The segment on the  left hand end, entirely falls during a non seizure event and hence is labelled as non-seizure. The  2nd and 3rd epochs from the left, either partially or fully occur during a seizure event and hence are labelled as seizure.   We apply to each epoch the wavelet transform.\par
\begin{figure}[ht]
    \centering
    \includegraphics[width=15cm, height= 8cm]{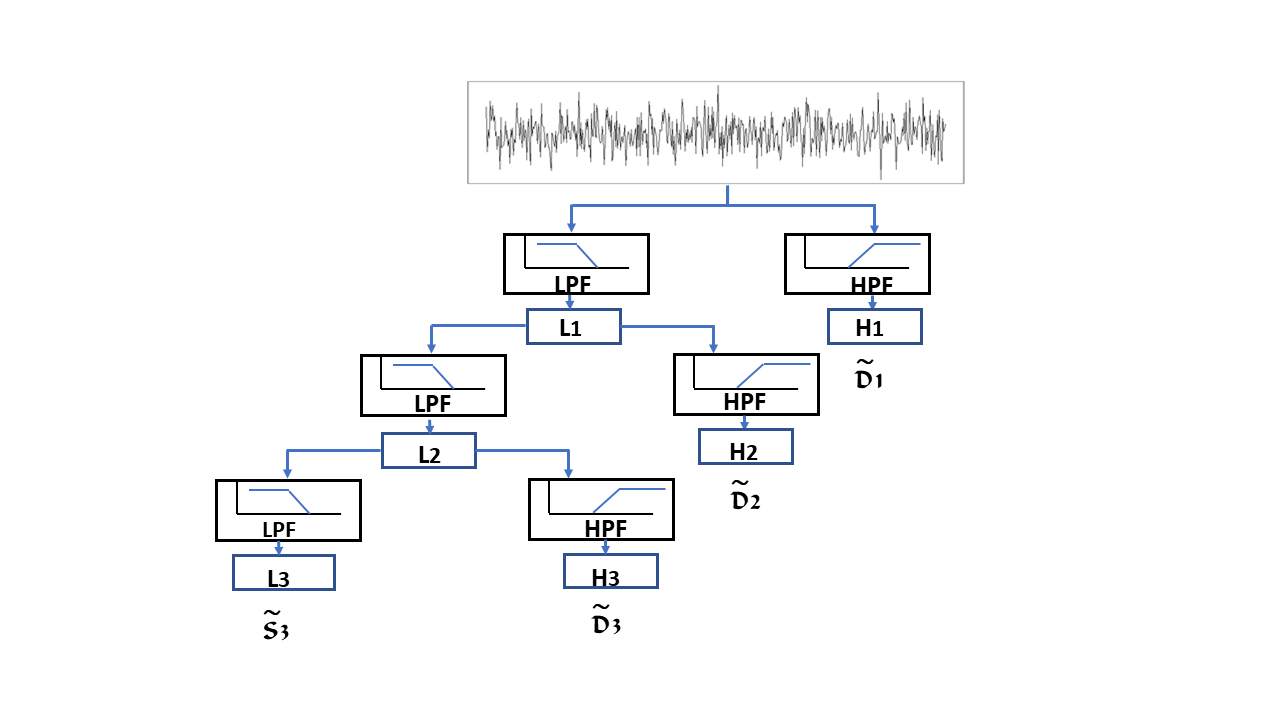}
    \caption{Wavelet Decomposition to 3 levels, \textit{from \cite{Gran:2019} } }
    \label{fig:Wavedecomp}
\end{figure}
\begin{figure}[ht]
    \centering
    \includegraphics[width=17cm, height=8cm]{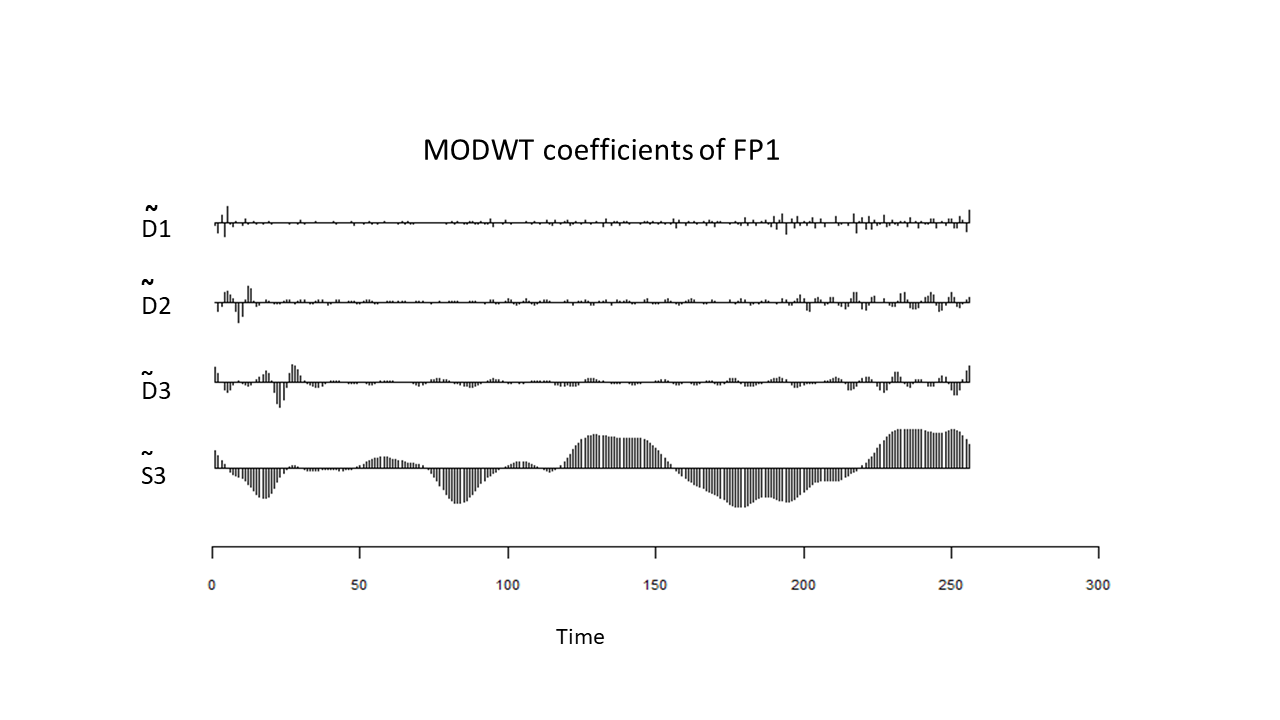}
    \caption{Wavelet Coefficients from 3 level decomposition, for an epoch}
    \label{fig:WavedDSlev}
\end{figure}

The decomposition function of the 
wavelet transform may be represented as a tree of low and high pass filters (LPF \& HPF),  with each step further decomposing the low pass filter, Figure \ref{fig:Wavedecomp} displays such decomposition to 3 levels. In this example the  resultant transform would consist of the levels $\{H1, H2, H3 \; \& \; L3\}$ and the usual nomenclature of the MODWT for these  levels  is $\{\widetilde{\mathcal{D}_1}, \widetilde{\mathcal{D}_2}, \widetilde{\mathcal{D}_3}\; \& \; \widetilde{\mathcal{S}_3} \}$ respectively. Each of these MODWT levels consists of wavelet coefficients, with $\{d_{i,j}\} \in \widetilde{\mathcal{D}_j}$ as detail coefficients and $\{s_{i,J}\} \in \widetilde{\mathcal{S}_J}$ for smooth coefficients\footnote{Similar to Section \ref{sec:RelW} for the DWT here $\widetilde{\mathcal{D}_j}$ and $\widetilde{\mathcal{S}_J}$ represent the MODWT levels $j$ or $J$ : $0 \le j \le J$}. Figure \ref{fig:WavedDSlev} shows the wavelet coefficients resulting from a 3 level MODWT, at each of the detail and smooth levels, across time for an epoch.\par
\begin{figure}[ht]
    \centering
    \includegraphics[width=15cm, height=7.5cm]{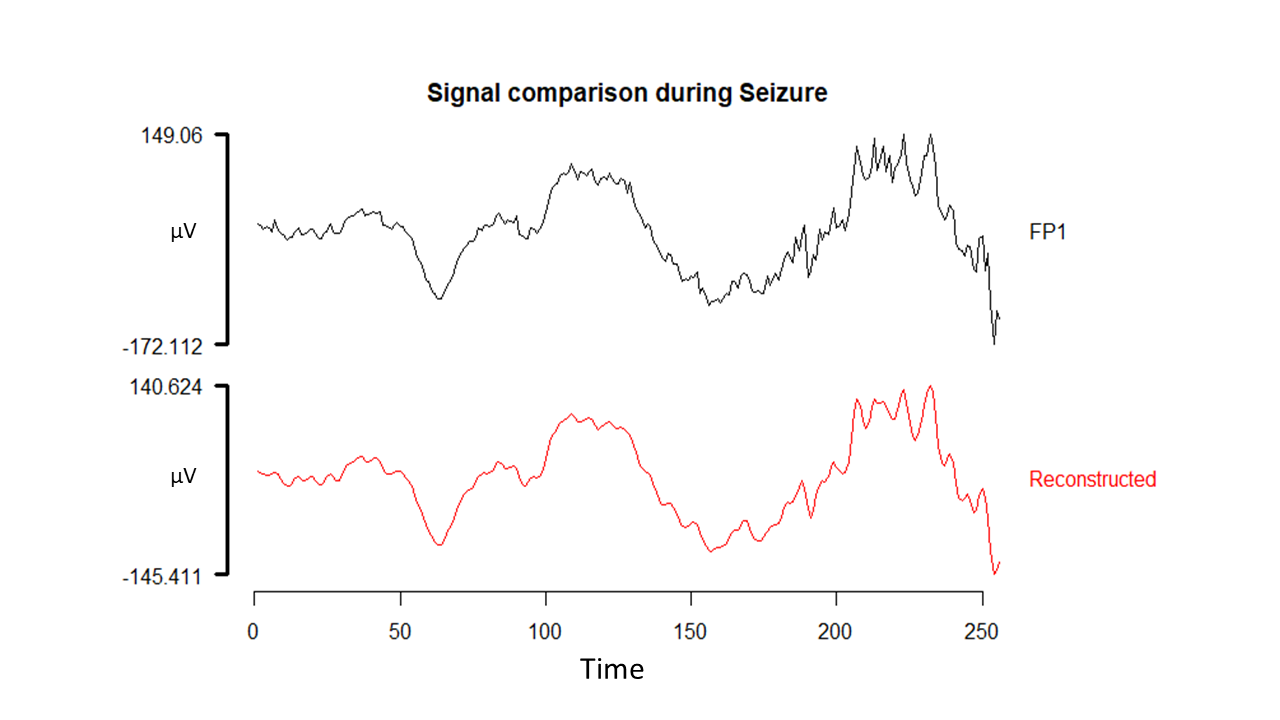}
    \caption{FP1 and  Reconstructed FP1, omitting level $\widetilde{\mathcal{D}_1}$, for an epoch}
    \label{fig:sigComp}
\end{figure}

EEG waves are conventionally grouped into frequency ranges and named: delta, theta, alpha, beta and gamma waves, where each set of these waves occupies a different non overlapping frequency bandwidth. Then dependent upon the sampling frequency of the initial data, we select our level of wavelet decomposition to align as close as possible with the frequency bandwidths required, to analyse the EEG signal \citep{Jacob:2018}.  %\textit{(If data was originally sampled at 256Hz then our decomposition would be as shown in  Table \ref{tab:freq} page \pageref{tab:freq}.)} 
However, some  frequency bands are unlikely to be related to the EEG event we seek to classify, as  it has been noted that EEG signals do not have any useful frequency components above 30 Hz \citep{sub:2007}. Therefore after wavelet decomposition, we omit or smooth the non-related frequency bands and reconstruct our signal using the inverse wavelet transform\footnote{The inverse wavelet transform is fully documented \citep[Chapter 5]{percival_walden_2000}.}. Figure \ref{fig:sigComp}   shows an original signal compared to the reconstructed signal where MODWT level $\widetilde{\mathcal{D}_1}$, has been omitted prior  to applying the inverse wavelet transform.  \par
%\subsubsection{feature extraction}
For each electrode \textit{(say, 1 to n)}, after decomposition and reduction of unwanted noise, the inverse transform produces a  reconstructed signal of same length as the original data.  We then take statistical parameters being: minimum, maximum, mean, standard deviation and normalised energy\footnote{Normalised energy ($NormE$) here is the sum of the squares of the time series values divided by the number of elements in the time series, i.e. Norm.Energy =  $ \frac{\sum_i^k X_i^2}{k}$} from each electrode's reconstructed signal and form the attributes. We have for each epoch; say epoch $i$\newline  $min_{i1}, max_{i1},mean_{i1}, std_{i1}, normE_{i1}, min_{i2}, max_{i2}, mean_{i2},\ldots, mean_{in}, std_{in}, NormE_{in}$. \,
We use the statistical parameters to  form a  list of attributes representing a single epoch. Table \ref{tab:ExAtt1} represents the data set produced in this step. The following steps add additional attributes to the data set as explained later. A data set is generated from the EEG signal of each patient in order to create patient specific training data set. One may also create a training data set combining all patients.  Algorithm \ref{Algo1} in Section \ref{sec:tech} outlines this Step 1 in a structured procedure format.

\begin{table}[ht]
\centering
\caption{Example: Data set resulting from Step 1}
\resizebox{\textwidth}{!}{%
 \begin{tabular}{||c| c c c c ||} 
 \hline
   & \multicolumn{4}{c||}{Attributes derived from each electrode's signal} \\
 $Individal_j$ & Elec-\textit{1} & Elec-\textit{2}  & ... & Elec-\textit{n}\\ 
 \hline\hline
 Epoch \texttt{1} &  $min_{11}, max_{11}, \ldots, normE_{11}$ &  $min_{12}, max_{12}, \ldots, normE_{12}$ & ...  & $min_{1n}, max_{1n}, \ldots, normE_{1n}$ \\ 
 Epoch \texttt{2}   &$min_{21}, max_{21}, \ldots, normE_{21}$ &  $min_{22}, max_{22}, \ldots, normE_{22}$ & ...  & $min_{2n}, max_{2n}, \ldots, normE_{2n}$\\
 \vdots & \vdots&   &  &\vdots \\
Epoch \texttt{m} & $min_{m1}, max_{m1}, \ldots, normE_{m1}$ &  $min_{m2}, max_{m2}, \ldots, normE_{m2}$ & ...  & $min_{mn}, max_{mn}, \ldots, normE_{mn}$ \\[1ex] 
 \hline
 \end{tabular}}
 \label{tab:ExAtt1}
\end{table}

\subsection{Step 2: Preparation of Attributes from Connectivity of Electrodes} \label{sub:WaveVar1}
Using the MODWT to decompose the time series from  each electrode per epoch, we derive wavelet variance, and similarly wavelet covariance and hence wavelet correlation between each of the transformed signals from each electrode.  Wavelet variance  is calculated at each level of wavelet decomposition  and is equal to the variance of the wavelet coefficients at that level. The sum of the wavelet variances from all decomposition levels equals the sample variance of the time series, similarly for wavelet covariance between each of the transformed signals. Covariance provides a measure of joint variation for two random sequences. The covariance of a random sequence with itself would be the variance.  see Appendix \ref{App:moDWT}, \ref{sub:wavecorr} and \cite[Chapter~8]{percival_walden_2000}. \par %From these values we could construct an adjacency matrix  highlighting the direct connections between the electrodes. 
Wavelet correlation  which is derived from wavelet covariance see Equation \ref{wave:corel}, provides us with a metric allowing us to determine association (joint variability) between the transformed signals from each electrode to another, during an epoch.  By setting a user defined lower bound on the correlation value between two electrodes, we may define if a connection (or association) exists between these electrodes. If the correlation value between two transformed signals is greater than or equal to the user defined lower bound\footnote{The value of this lower bound may be derived by the number of connections available  determined by that specific value, given that we need to derive statistical parameters from these connection values.} then we conclude that a connection exists between the two during the epoch. \par
We then represent those connections in  an  adjacency matrix where a connection is represented by the value 1. See Table \ref{tab:ExAdl} as an example of an adjacency matrix.  Here, Electrode 1 (Elec-\textit{1}) and Electrode 2 (Elec-\textit{2}) are connected since the correlation between the signals, as calculated by Equation \ref{wave:corel}  is higher than the user defined lower bound. From the derived  adjacency matrix we take the sum of the rows.\newline  In Table \ref{tab:ExAdl},
Electrode 1  is directly connected to two other electrodes i.e Elec-\textit{2} \& Elec-\textit{3}, therefore sum of the row is 2. That is the degree of connection is equal to 2.\newline
Electrode 3 is directly connected to five  other electrodes, therefore the sum of row is 5. Hence the degree of connection is equal to  5. \newline
For each electrode in our network we now have a degree of connection or number of direct connections during the epoch. Figure \ref{fig:WaveConn} is a representative example of this.
From these row sums (\textit{or degree of connection})  we derive the statistical parameters: mean, maximum, 1st Quartile, 3rd Quartile and skewness. We represent these as: $meanC, maxC, \ldots, skewC$. This provides us with additional attributes for the  epoch in question. These additional attributes for each epoch are combined  with the data set outlined in Step 1 (as shown in Table \ref{tab:ExAtt1}) increasing its dimension. The updated data set is shown in  Table \ref{tab:ExAtt2a}. A procedural  format of this Step 2 is shown in Algorithm \ref{Algo1}.

\begin{table}[ht] 
\centering

\caption{Example: Adjacency matrix }

\resizebox{\textwidth}{!}{%
 \begin{tabular}{||c c c c c c c c c c|| c ||} 
 \hline
  & Elec-\textit{1} & Elec-\textit{2} & Elec-\textit{3} & Elec-\textit{4} & Elec-\textit{5} & Elec-\textit{6} & Elec-\textit{7} &\ldots& Elec-\textit{n} &\textit{Sum}\\ [0.2ex] 
 \hline\hline
 Elec-\textit{1} & 0 & 1 & 1 & 0 & 0 & 0 & 0 &\ldots & 0  &\textit{2}\\ 
 \hline
 Elec-\textit{2} & 1 & 0 & 1 & 0 & 0 & 1 & 1 & &0 &\textit{4}\\
 \hline
 Elec-\textit{3} & 1 & 1 & 0 & 1 & 1& 1 & 0 & &0&\textit{5}\\
 \hline
 Elec-\textit{4} & 0 & 0 & 1 & 0 & 0 & 0 & 0 & &0 &\textit{1}\\
 \hline
 Elec-\textit{5} & 0 & 0 & 1 & 0 & 0 & 0 & 0&\ldots &0&\textit{1}\\  
 \hline
 Elec-\textit{6} & 0 & 1 & 1 & 0 & 0& 0 & 1 & &0&\textit{3}\\  
 \hline
 Elec-\textit{7} & 0 & 1 & 0 & 0 & 0 & 1 & 0& &0&\textit{2}\\  
 \hline
 \vdots  & &\vdots & &  & & & \vdots & &&\vdots\\
 \hline
  Elec-\textit{n} & 0 & 0 & 0 & 0 & 0 & 0 & 0&\ldots &0&\textit{0}\\  
 \hline
\end{tabular}}
%\end{center}
\label{tab:ExAdl}
\end{table}

\begin{table}[ht]
\centering
\caption{Example: Data set resulting after Step 2}
\resizebox{\textwidth}{!}{%
 \begin{tabular}{||c| c c c c ||} 
 \hline
   & \multicolumn{4}{c||}{Attributes derived by Step 1  and Step 2} \\
   & \multicolumn{3}{c}{Step 1} & \multicolumn{1}{c||}{Step 2}\\
 $Individal_j$ & Elec-\textit{1}& ...  & Elec-\textit{n}   & Connectivity \\ 
 \hline\hline
 Epoch \texttt{1} &  $min_{11}, max_{11}, \ldots, normE_{11}$ &...  &$min_{1n}, max_{1n}, \ldots, normE_{1n}$,  & $meanC_1, maxC_1, \ldots, skewC_1$ \\ 
 Epoch \texttt{2}   &$min_{21}, max_{21}, \ldots, normE_{21}$ &...  &$min_{2n}, max_{2n}, \ldots, normE_{2n}$, & $meanC_2, maxC_2, \ldots, skewC_2$\\
 \vdots & \vdots&   & \vdots  &\vdots \\
Epoch \texttt{m} & $min_{m1}, max_{m1}, \ldots, normE_{m1}$ &...  &$min_{mn}, max_{mn}, \ldots, normE_{mn}$,  & $meanC_m, maxC_m, \ldots, skewC_m$ \\[1ex] 
 \hline
 \end{tabular}}
 \label{tab:ExAtt2a}
\end{table}

% \begin{table}[h]
% \centering
% \caption{Example: Data set resulting from Step 2.}
%  \begin{tabular}{||c| c   ||} 
%  \hline
%  $Individal_j$ & Attributes derived from degree of connection \\ 
%  \hline\hline
%  Epoch \texttt{1} &  mean \texttt{1}, max \texttt{1}, skewness \texttt{1} \\ 
%   Epoch \texttt{2} &mean \texttt{2}, max \texttt{2}, skewness \texttt{2}   \\
%  \vdots & \vdots  \\
% Epoch \texttt{n} & mean \texttt{n}, max \texttt{n}, skewness \texttt{n} \\ [1ex] 
%  \hline
%  \end{tabular}
%  \label{tab:ExAtt2}
% \end{table}

%\subsubsection{feature extraction}

%The data initially  chosen consisted of results from 23 electrodes or channels,  as one channel was the reverse of the other, ie channel 3: T7-P7, channel 19:  P7-T7, we dropped channel 19 when calculating the wavelet variance and hence from the adjacency matrix construction, had 22 row sums to derive the statistical parameters from.

\subsection{Step 3. Preparation of Attributes from Global Efficiency of Electrodes.}\label{sub:eleceff}
\begin{figure}[ht]
    \centering
    \includegraphics[height = 5.5cm, width = 14cm]{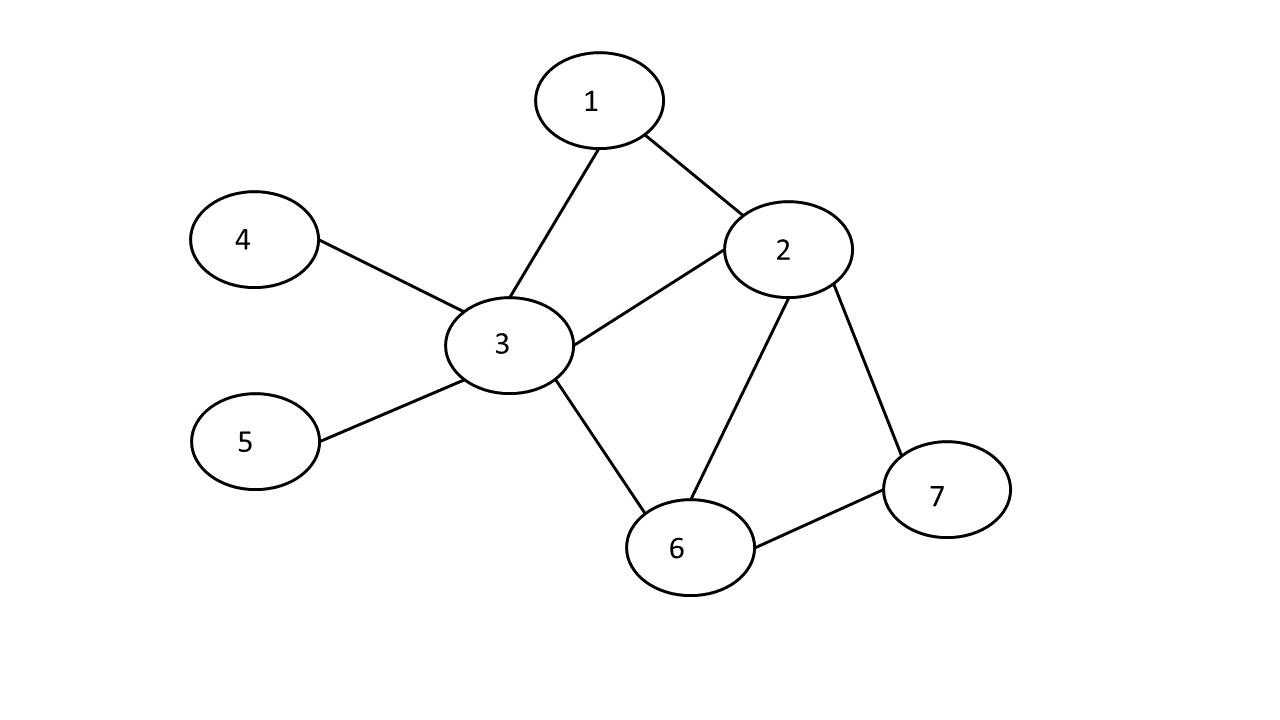}
    \caption{ A Visual  representation of Table \ref{tab:ExAdl}}
    \label{fig:netEXl1}
\end{figure}

 Section \ref{sub:WaveVar1} shows that from  the wavelet coefficients  we have a  methodology for the  construction of an adjacency matrix derived from the correlation between the electrodes.  From that adjacency matrix we are able to  construct a visual representation of the network  or Graph. Figure \ref{fig:netEXl1} shows a sub-graph derived from only the first seven electrodes of Table \ref{tab:ExAdl}.  In Figure \ref{fig:netEXl1}  an electrode is represented as a numbered node and a pair of electrodes/nodes having a correlation above the threshold  are connected through an edge between them.
 Figure \ref{fig:netEXl1}, has seven nodes/vertices \textit{(i.e. N = 7)} and nine edges. Each single edge has an assumed unit length of 1. From this constructed network diagram (\textit{and hence the adjacency matrix}) we derive how “efficiently”\footnote{Efficiency of $\text{node}_i$  is related  to the sum of the inverses of the minimum distances between $\text{node}_i$ and all other nodes in the network.} the electrodes are communicating across their network of connections. We calculate the global efficiency ($Eglob_i$) for each node. \par 
 
 For every electrode $i$ we compute  the reciprocal of  shortest path length ($R_{i,j}$)  to each other electrode $j$.  For example the shortest path length from electrode 5 to Electrode 7 is three, hence $R_{5,7}$ is $\frac{1}{3}$. This is shown in the corresponding cell (encircled) in Table \ref{tab:ExGlobC1}. For each electrode $i$ we then then compute $ \frac{\sum \limits_{j=1}^N R_{i,j}}{N-1}$ where $N$ is the number of nodes and $R_{i,i} = 0$. The last row of Table \ref{tab:ExGlobC1} shows these values. We call these values global efficiency ($Eglob_i$) values of an $\text{electrode}_i$ for an epoch, see definition \ref{def:globeff}.
%  We demonstrate this for Figure \ref{fig:netEXl1},
% where Table \ref{tab:ExGlobC1} displays the reciprocal of the minimum path length between Electrodes/nodes, with the final row as the sum of each column divided by \textit{N-1}, which follows from Definition \ref{def:globeff} Appendix \ref{Defs}. In this example, to calculate an entry in Table \ref{tab:ExGlobC1}; being Elec5 to Elec7.  As the minimum number of edges (or paths) transversed  from Electrode 5 to Electrode 7 is three, hence reciprocal of that minimum path length = 1/3. We then do exactly the same for Electrode 5 to each other Electrode. Sum the column  and divide by \textit{N-1}(i.e = 6).   % The final in the table final row, labelled Sum/N-1) which is the sum of respective column, divided by \textit{N-1}(i.e = 6).
% The final row in Table \ref{tab:ExGlobC1} represents the global efficiency ($Eglob_j$) for each electrode/node  where $Eglob_j : 0 \le  Eglob_j \le 1 ,\, j \in \{1,7\}$, in this example. 

\begin{table} [ht]
\centering

\renewcommand{\arraystretch}{1.6}

\caption{Example: \textbf{Global Efficiency}}
 \begin{tabular}{||c c c c c c c c||} 
 \hline
  & Elec1 & Elec2 & Elec3 & Elec4 & Elec5 & Elec6 & Elec7\\ [0.2ex] 
 \hline\hline
 Elec1 & 0 & 1 & 1 & 1/2 & 1/2 & 1/2 & 1/2 \\ 
 \hline
 Elec2 & 1 & 0 & 1 & 1/2 & 1/2  & 1 & 1 \\
 \hline
 Elec3 & 1 & 1 & 0 & 1 & 1 & 1 & 1/2\\
 \hline
 Elec4 & 1/2 & 1/2 & 1 & 0 & 1/2 & 1/2 & 1/3 \\
 \hline
 Elec5 & 1/2 & 1/2 & 1 & 1/2 & 0 & 1/2 & \circled{1/3}\\  
 \hline
 Elec6 & 1/2 & 1 & 1 & 1/2 & 1/2 & 0 & 1 \\  
 \hline
 Elec7 & 1/2 & 1 & 1/2 & 1/3 & 1/3 & 1 & 0\\  
 \hline
 \textit{Sum/(N-1)}  & \textbf{2/3} & \textbf{5/6} & \textbf{11/12} & \textbf{5/9} & \textbf{5/9} & \textbf{ 3/4} & \textbf{11/18}\\
 \hline
\end{tabular}
%\end{center}
\label{tab:ExGlobC1}
\end{table}

 %{\color{red}Up to here}
%\subsubsection{feature extraction}
%This efficiency value  represents how effectively an electrode communicates or transfers information within the  network. %Table \ref{tab:ExGlobC1} displays  From this sum of the minimum path lengths for by
From these global efficiency values for an epoch as shown in the last row of Table \ref{tab:ExGlobC1} we derive statistical parameters:  mean, maximum, 1st Quartile, 3rd Quartile \&  skewness. We represent them as $MeanG_i, MaxG_i, \dots. SkewG_i$ for the $i$th epoch. These attributes are again added to the data set as shown in Table \ref{tab:ExAtt3a}.  Again this step in procedural form is seen in Algorithm \ref{Algo1}. %and place directly following the attributes derived in \textbf{Step 1} and \textbf{Step 2} of our method.
% \textit{These three steps combined together provides  a total 115 + 5 + 5 = 125  attributes.  Each step is further outlined in Algorithm\ref{Algo1}.} 

\begin{table}[ht]
\centering
\caption{Example: Data set resulting after Step 3.}
\resizebox{\textwidth}{!}{%
 \begin{tabular}{||c| c c c c ||} 
 \hline
   & \multicolumn{4}{c||}{Attributes derived by Step 1, Step 2 \& Step 3} \\
  & Step 1 &  & Step 2 &  Step 3 \\
 $Individal_j$ & Elec-\textit{1 to n}&   & Connectivity   & Global Efficiency \\ 
 \hline\hline
 Epoch \texttt{1} &  $min_{11}, max_{11}, \ldots, normE_{1n}$, &  &$meanC_1, maxC_1, \ldots, skewC_1$,  & $MeanG_1, MaxG_1, \dots SkewG_1$ \\ 
 Epoch \texttt{2}   &$min_{21}, max_{21}, \ldots, normE_{2n}$, &  &$meanC_2, maxC_2, \ldots, skewC_2$, & $MeanG_2, MaxG_2, \dots SkewG_2$\\
 \vdots & \vdots&   & \vdots  &\vdots \\
Epoch \texttt{m} & $min_{m1}, max_{m1}, \ldots, normE_{mn}$, &  & $meanC_m, maxC_m, \ldots, skewC_m$,  &  $MeanG_m, MaxG_m, \dots SkewG_m$ \\[1ex] 
 \hline
 \end{tabular}}
 \label{tab:ExAtt3a}
\end{table}

\begin{table}[ht]
\centering
\caption{Example: Data set resulting from the 3 Steps and attaching label}
\resizebox{\textwidth}{!}{%
 \begin{tabular}{||c| c c c l||} 
 \hline
   & \multicolumn{4}{c||}{Attributes derived from each Step } \\
 $Individual_j$ & \textit{Step 1}  &\textit{Step 2}   & \textit{Step 3}& Label\\ 
 \hline\hline
 Epoch \texttt{1} &  $min_{11}, max_{11},\ldots, normE_{1n}$,& $meanC_1, maxC_1, \ldots, skewC_1$, & $MeanG_1, MaxG_1, \dots SkewG_1$,& non-seizure\\ 
 Epoch \texttt{2}   &$min_{21}, max_{21},\ldots, normE_{2n}$,& $meanC_2, maxC_2, \ldots, skewC_2$, & $MeanG_2, MaxG_2, \dots SkewG_2$, &seizure\\
 \vdots & \vdots& \vdots   &\vdots  & $\qquad$ \vdots\\
Epoch \texttt{m} &  $min_{m1}, max_{m1},\ldots, normE_{mn}$,& $meanC_m, maxC_m, \ldots, skewC_m$,  &$MeanG_m, MaxG_m, \dots SkewG_m$, &non-seizure \\[1ex] 
 \hline
 \end{tabular}}
 \label{tab:ExAttall}
\end{table}

  \subsection{Step 4. Assigning Class labels and Training Classifiers} \label{subsec:classM}
% From the results  from our first three steps,  we construct a data set from the transformed signal data. As an example see Table \ref{tab:ExAttall}, where we have transformed the multivariate EGG data into a univariate list of attributes per Epoch, per individual. This is achieved by arranging  the derived statistical parameters resulting from each of the first three steps of our proposed method into a single data string  as well as attaching a class label. This class label is determined  from the original signal data, as per Figure \ref{fig:EEGdata}, where if any segment of the epoch is enclosed within the period identified as seizure, then the epoch is labelled seizure, otherwise non-seizure.  We  apply a cost sensitive / class imbalanced  classifier to our transformed data. Here we  use an individual's incoming data, per epoch  as our testing data. For training data we build a  model on data, transformed via our method, which at this stage would not include any data from the individual's data being tested.
In this step we first assign the class labels to the epochs. That is for every row/epoch in Table \ref{tab:ExAtt3a} we assign an additional column with a class label, see Table \ref{tab:ExAttall}. If the epoch entirely falls within a non-seizure period then the epoch is labelled as non-seizure (as also explained in Section \ref{sub:step1} and Figure \ref{fig:EEGdata})  otherwise the epoch is labelled as seizure. As the consequence of misclassification of a seizure record as a non-seizure record can be much higher than a non-seizure record with a seizure prediction, we use a cost sensitive classifier. Thus we aim to reduce the misclassification of seizure records with non-seizure records (i.e. False Negative predictions). Once we train our cost sensitive classifier on the training data set, being  Table \ref{tab:ExAttall}, we are ready to predict/detect seizure in the new unlabelled signal. When we receive new unlabelled signal we first convert into a record as shown in Table \ref{tab:ExAtt3a}. The trained classifier is then used to label the record.

\begin{algorithm}[ht!]
\SetAlgoLined
\SetKwInOut{Input}{input}\SetKwInOut{Output}{output}
\Input{ A Signal input $DB_I$,  of individuals' EEG recordings}
\Output{$A_{i,j}$ : Output for individual  per epoch from the Cost sensitive Classifier.}
%\KwData{EGG Data}
%\KwResult{Output from Classifier }
Initialization: Build transformed database\\
$I_j \leftarrow DB_I$ //Download initial data file per individual\\

\For{All individuals $I_j$} {
$I_{i,j} \leftarrow I_j$ //Segment individual's data into 1 second Epochs\\
\For{Each Individual by Epoch $I_{i,j}$}{
 \textbf{Step 1.}\\
\For{Each Electrode $I_{i,j,e}$}{
$\mathbf{\widetilde{W}_i} \leftarrow I_{i,j,e}$ // Apply MODWT, decompose to required  levels \\
 $I_{i,j,e,R} \leftarrow \mathbf{\widetilde{W}_i}$// Omit noise levels, use Inverse MODWT to reconstruct reduced signal \\
 $Sp_i \leftarrow I_{i,j,e,R,}$ // Derive statistical parameters from new signal\\
}
$\text{RW}_{i,j,e} \leftarrow Sp_1,Sp_2, \dots, Sp_n $ //Form attributes from statistical parameters \\

}
\For{Each Individual by epoch $I_{i,j}$}{
\textbf{Step 2.} \\
\For{Each Electrode $I_{i,j,e}$}{
$\mathbf{\widetilde{W}_i} \leftarrow I_{i,j,e}$ // Apply MODWT, decompose to first wavelet detail level \\
%$ \nu_X^2(\mathcal{T}_i) \leftarrow \mathbf{\widetilde{W}_i}$ // Derive wavelet variance\\
}
\For{Each electrode pair $\{I_{i,j,\tau},I_{i,j,\zeta}\}$ }{
$cov \{\widetilde{W}_{i,\tau},\widetilde{W}_{i,\zeta}\} \leftarrow \widetilde{W}_{i,\tau}, \widetilde{W}_{i,\zeta}$ // derive wavelet covariance between electrodes\\
$\rho_{\tau \zeta} \leftarrow cov\{\widetilde{W}_{i,\tau},\widetilde{W}_{i,\zeta}\} $//derive correlation from standardised covariance\\
}
$Thresh\leftarrow 0 < t_h \le 1$ // determine correlation threshold level $t_h$\\
$Adj_{ee} \leftarrow 1 \text{where} \, \rho \ge Thresh$ //construct adjacency matrix \\
%$\sum\limits_{i=1}^er_i : r \text{ is a row} \in  Adj_{ee} $// calculate row sums of adjacency matrix\\
$Sp_i \leftarrow \sum r_{i}$ //derive statistical parameters from row sums  of adj matrix\\
$C_{i,j} \leftarrow Sp_1,Sp_2, \dots , Sp_k $ // Use statistical parameters to form attributes\\
$\text{RW}_{i,j,e} , C_{i,j}  \leftarrow C_{i,j}$ // combine with attributes from Step 1. \\
\textbf{Step 3.} \\

$ Eglob_i \leftarrow Adj_{ee} $ //From Adjacency matrix derive Global efficiency \\

$ Sp_i \leftarrow Eglob_i$ // derive statistical parameters\\
$G_{i,j} \leftarrow Sp_1,Sp_2, \dots , Sp_k $ // Use statistical parameters to form attributes\\
$\text{RW}_{i,j,e}, C_{i,j}, G_{i,j} \leftarrow G_{i,j}$ // combine with attributes from Step 2\\

    }

  \textbf{Step4.} $DB_T \leftarrow \text{RW}_{i,j,e}, C_{i,j}, G_{i,j}, label$ // add label and build database \textit{(for training classifier upon)} \\
 %$A_j \leftarrow DB $ //Apply CSForest via leave one out cross validation
  }  
  %Step 4. \\
     $A_{i,j} \leftarrow DB_T $ //Apply Cost sensitive Classifier  $\qquad \blacksquare$ \\

\caption{EEG Analysis via MODWT decomposition and Wavelet Correlation}\label{Algo1}
\end{algorithm}

 %{\color{red}{\textit{continue}} HERE!!} 
\section{Experimental Results} \label{sec:tech}

\subsection{Data}
We used data from \cite{PhysioNet}, which was collected at the Children's Hospital Boston,
Massachusetts (MIT). This consists of EEG recordings  from paediatric subjects with intractable seizures. There are recordings from 24 different investigations (cases)  from 23 subjects (5 males, 17 females and 1 with gender not specified). The International 10-20 system of EEG electrode positions (\textit{for 23 electrodes}) and nomenclature was used for these recordings with the data sampled at 256Hz. More details about the data set may be found from  https://archive.physionet.org/pn6/chbmit/. %{\color{red} UP TO HERE} \par 
For this study we used a single file from each of 12 individuals, these individuals were identified in the MIT data by the case numbers; 1, 2, 3, 4, 5, 6, 7, 8, 10, 11, 23 and 24. 
 The cases were selected on the basis of
 
\begin{itemize}
    \item the data across cases was reported  with same identical contiguous electrode sequence,\newline i.e.$\{ FP1, F7, T7,\ldots, FT10, P8 \}$.
    \item each case was a different individual
    \item limited selection to files up to two hours in duration 
    \item no seizure event within the first 10 minutes.
\end{itemize}    
    Most of the selected files were approx one hour in duration. As the files have 256 data points per second for each electrode signal, therefore one hour equates to $3600 \times 256 = 921600$ data points, per electrode.\par
A few files were two hours in length,  for these files we selected either the first or second hour depending upon which hour had the first seizure event. This raw data is presented with indicated periods of seizure, we subset the data  into one second intervals (i.e. an epoch length of one second)  and attached labels indicating the class of that epoch, i.e. non-seizure or seizure.  We have approx 1 hour of recording from each patient. For two of the patients there were no seizures recorded within the hour sample, we still included these two into our training set. Of these two, one patient simply had no seizures recorded in files of up to two hours duration, the other patient's files had the sequence in which the  electrodes were reported, altered before first file with a seizure occurred.

\subsection{ Synopsis of  results and comparisons } \label{sub:results}
In this study True Positive (TP) refers to events  where epochs labelled  seizure were classified as seizure. False Negative  (FN) refers to events where seizure labelled epochs were classed as non-seizure.
True negative (TN) refers to epochs labelled non-seizure and classified non-seizure and False Positive (FP) refers to a an epoch labelled non-seizure and classified seizure.
 We use the following metrics or indicators: Accuracy, Recall, Precision and f-score. They are defined as,

\begin{eqnarray*}
Accuracy\,\% & =  &\frac{TP + TN}{TP + FN + TN + FP } \times 100\\
Recall  & = & \frac{TP}{TP + FN}\\
Precision & = & \frac{TP}{TP+FP}\\
f-score & = & \frac{2}{(Recall)^{-1} + (Precision)^{-1}} =  2 \times \frac{Precision \times Recall}{Precision + Recall}
\end{eqnarray*}

We use  these indicators to compare  our methodology to the other methods.
As we are interested in correctly  classifying seizure  we  use a cost sensitive/class imbalanced algorithm. Here we choose CSForest \citep{sier:2014} and set the cost penalty structure as shown in Table \ref{tab:CScost}.

\begin{table}[ht]
\centering
 \caption{ Penalty Cost Structure for CSForest. }
\begin{tabular}{l|l|c|c|c}
\multicolumn{2}{c}{}&\multicolumn{2}{c}{Actual Value}&\\
\cline{3-4}
\multicolumn{2}{c|}{}&Positive&Negative&\multicolumn{1}{c}{}\\
\cline{2-4}
\multicolumn{1}{c|}{Predicted}& Positive & $TP=1$ & $FP = 1$ & \\[0.2ex]
\cline{2-4}
\multicolumn{1}{c|}{Value} & Negative & $FN = 15$ & $TN = 0$ & \\[0.2ex]
\cline{2-4}

\end{tabular}

 %\caption{ Penalty Cost Structure for CSForest. }
\label{tab:CScost}

\end{table}

%\begin{itemize}
 %   \item  True positive  = 1,  True negative  = 0
 %   \item  False positive  = 1,  False negative  = 15,  \textit {where seizure is defined as a positive event.}

%\end{itemize}
We arrived at this cost structure by training on all individuals in the transformed data set, except individual 1. From this initial reduced data set, we implemented  a 66\% training 34\% test split on this data set for validation of the CSForest model (i.e we repeated this training using  66\% of data for training  and testing on 34\%, altering the penalty cost, with the objective of increasing the f-score while maintaining a high Recall value).  As this penalty cost structure shown in Table \ref{tab:CScost} provided us with a close to optimal result, in regards to f-score\footnote{The f-score is the harmonic mean of precision and recall, sometimes called  a  balanced f-score.},  we left the CSForest  cost structure unchanged throughout the rest of  experiment across all individuals in our data set.\par

%Table \ref{table:comp} displays the results for each individual (\textit{In.}), which had a seizure state.
We also compared our results to the two other methods outlined in Section \ref{sub:CopM}  and  used leave-one-out cross validation for all methods. By using the macro average method, see Definition \ref{def:maf} Appendix \ref{Defs}, for determining an average f-score across individuals per method,  our method  returned a higher average f-score than the other two comparative methods, as shown in Table \ref{table:fscore}. %\textbf{1} being from \cite{sidd:2018}  and method \textbf{2} from \cite{Chen:2017} respectively. 
For the f-score results at an individual level see Table \ref{table:compA}. Our technique (labelled Method No.\textbf{3} in Tables \ref{table:fscore} \& \ref{table:compA}) usually returns better results in  respect to this balanced f-score across these individuals. We choose the f-score as it weights both Precision and Recall equally. \par 

%There were two other individuals included in our data set and  neither of these had a seizure state in the file selected. Since there is no seizure record, all three classification methods all provided near perfect results in regards to accuracy and f-score for these two individuals, (\textit{not shown}).

\begin{table}[ht]
\caption{Average f-score results via Method.}
\label{table:fscore}
\begin{center}
 \begin{tabular}{||c c c ||} 
 \hline
 Method No. & Method origin  & Average f-score  \\ [.2ex] 
 \hline\hline
 \rule{0pt}{12pt}{1} & Siddiqui et al. (2018) &0.37  \\ [.2ex]
 \hline 
 \rule{0pt}{12pt}{2} & Chen et al. (2017) & 0.42   \\ [0.2ex] 
 \hline
  \rule{0pt}{12pt}\textbf{3} & Our Proposed Technique & \textbf{0.58}   \\ [0.2ex]
 
 \hline

 \hline
\end{tabular}
\end{center}
\end{table}

\begin{table}[ht]
\caption{f-scores via Individual across Methods}
\label{table:compA}
\begin{center}
%\resizebox{\textwidth}{!}{%
\begin{tabular}{ |c||c|c|c||c|||c|c|c|  }
 \hline
 \multicolumn{8}{|c|}{\bfseries Comparative results} \\
 %\multicolumn{10}{|c|}{\bfseries method numbers \hspace{3.4cm} method numbers}\\
 \multicolumn{8}{|l|} {\hspace{1.9cm}\textbf{Method Number}\hspace{3cm}\textbf{Method Number}} \\
  \multicolumn{8}{|l|} {\hspace{3.1cm}{f-scores }\hspace{4.85cm}{f-scores}} \\
% \multicolumn{5}{l}{\hspace{3.9cm}\textbf{Method Number}}&\multicolumn{5}{l}{\hspace{3.9cm}\textbf{Method Number}}\\
 \hline
 \textit{Ind. No.}&  \textbf{1} & \textbf{2} & \textbf{3} &\textit{Ind. No.}& \textbf{1} & \textbf{2} & \textbf{3}\\
 \hline
  \textit{1}
    &0.136 & 0.178&  \textbf{0.646} & \textit{5}  &0.664 & 0.550 & \textbf{0.735} \\
    \hline
 \textit{2} 
   &0.081 & 0.062& \textbf{0.326}& \textit{7} &0.328 & 0.724 & \textbf{0.840}\\
   \hline
 \textit{3}
   &0.213 & 0.042& \textbf{0.385} & \textit{8} &0.32 & 0.037 & \textbf{0.455}\\
   \hline
   \textit{4} 
      &0.039 & 0.0&  \textbf{0.185} & \textit{10} &0.556 & 0.87 & \textbf{0.881}\\
     \hline
   \textit{23} 
      &0.018 & 0.0&  \textbf{0.395} &\textit{24}  &0.562 & 0.584 & \textbf{0.574}\\
  
 \hline
\end{tabular}
\end{center}

\end{table}

\begin{remark}There were two other individuals included in our data set and  neither of these had a seizure state in the file selected. Since there is no seizure record, all three classification methods all provided near perfect results in regards to accuracy and f-score for these two individuals, (\textit{not shown}).
\end{remark}

\subsubsection{ Metrics chosen }   In our situation we wish maximise True Positives, i.e. those with seizure classified as seizure as well as minimise False Negatives, i.e. a seizure classified as a non-seizure. The metric Recall is important here. Recall provides a measure of the True Positives,  divided by the sum of  True Positives and False Negatives. 
Alternatively, Precision which is equal to  True positives divided by  the sum of True Positives and False Positives. Here False Positives are those in a non-seizure state but classified as seizure. Therefore while we aim to detect all seizures as seizures, classifying non-seizures as seizures is at worst, a waste of resources. However classifying seizures as non-seizures may lead to catastrophic outcomes.

\begin{table}[ht]
\caption{Statistical results via Individual across Methods}
\label{table:comp}
\begin{center}
\resizebox{\textwidth}{!}{%
\begin{tabular}{ |c|p{1.72cm}||c|c|c||c|p{1.72cm}||c|c|c|  }
 \hline
 \multicolumn{10}{|c|}{\bfseries Comparative results} \\
 %\multicolumn{10}{|c|}{\bfseries method numbers \hspace{3.4cm} method numbers}\\
 \multicolumn{10}{|l|} {\hspace{4.2cm}\textbf{Method Number}\hspace{5cm}\textbf{Method Number}} \\
% \multicolumn{5}{l}{\hspace{3.9cm}\textbf{Method Number}}&\multicolumn{5}{l}{\hspace{3.9cm}\textbf{Method Number}}\\
 \hline
 \textit{Ind. No.}& Metric& \textbf{1} & \textbf{2} & \textbf{3} &\textit{Ind. No.}& Metric & \textbf{1} & \textbf{2} & \textbf{3}\\
 \hline
  1 &Accuracy\%& 98.94    & 98.87&   99.03 & 5 & Accuracy\%& 97.47&98.00& 98.56\\
      &Recall&   0.073  & 0.098   &0.780&     &Recall&  0.776  &0 .379   &0.621\\
     &Precision &1 & 1&  0.552&              &Precision &0.581 & 1  &  0.900\\
    & f-score   &0.136 & 0.178&  \textbf{0.646} &   & f-score   &0.664 & 0.550 & \textbf{0.735} \\
    \hline
 2 &Accuracy\% & 74.77    &97.46&   94.35 & 7 & Accuracy\%& 96.48&98.87& 99.28 \\
    &Recall&   0.488  & 0.037  &0.598 & &Recall&  0.330  & 0.567   &0.732\\
    &Precision &0.044 & 0.200&  0.224&  &Precision &0.327 & 1&  0.986\\
   & f-score   &0.081 & 0.062& \textbf{0.326}&  & f-score   &0.328 & 0.724 & \textbf{0.840}\\
   \hline
 3 &Accuracy\% & 97.94    &98.72&   96.89 & 8 & Accuracy\%& 95.39&95.61& 92.08 \\
    &Recall&   0.208  & 0.021   &0.729& &Recall&   0.241  & 0.018   &0.735\\
    &Precision &0.217 & 1&  0.261&  &Precision &0.476 & 1&  0.33\\
   & f-score   &0.213 & 0.042& \textbf{0.385} &  & f-score   &0.32 & 0.037 & \textbf{0.455}\\
   \hline
   4 &Accuracy\% & 98.67    &98.36&  94.83 & 10 & Accuracy\%& 97.77&99.61& 99.61 \\
      &Recall&   0.02  &0.0   &0.429& &Recall&   0.82  & 0.77  &0.813\\
      &Precision &0.50 & 0.0&  0.118&  &Precision &0.420 & 1 &  0.963\\
      & f-score   &0.039 & 0.0&  \textbf{0.185} &  & f-score   &0.556 & 0.87 & \textbf{0.881}\\
     \hline
   23 &Accuracy\% & 96.89    &96.86&   96.25 & 24 & Accuracy\%& 98.91&98.97& 98.81 \\
      &Recall&   0.009 & 0.0   &0.389& &Recall&   0.50  & 0.520   &0.580\\
      &Precision &1 & 0.0&  0.40&  &Precision &0.641 & 0.667&  0.569\\
      & f-score   &0.018 & 0.0&  \textbf{0.395} &  & f-score   &0.562 & 0.584 & \textbf{0.574}\\
  
 \hline
\end{tabular}}
\end{center}
\end{table}

Hence we consider Recall slightly more relevant than Precision for this exercise as Precision does not take into account False Negatives. Precision does however consider False Positives.  Table \ref{table:comp} which highlights all the chosen indicators at an individual level, demonstrates that our proposed method \textbf{(Method Number 3)}, returns higher values of Recall across nearly all individuals. 
In some instances Precision values from the comparative methods are higher than from our method, i.e. Individual No 1.  This could simply be  a situation  of a small proportion of the overall number of  seizures classified as seizures but with no non-seizures classified as seizures. There is no indication in this Precision  metric of how many False Negatives were generated in the classification process. This is why we have chosen f-score as our main indicator of overall performance, to balance out Recall and Precision. Accuracy, shown in Tables \ref{table:comp} \& \ref{table:comp2a} is a misleading metric for such imbalanced data. In our final data set of all individuals we have approx 42816 records of which 822 are labelled seizure. i.e. 1.92\%. Hence if we simply classify all records as non-seizure then our Accuracy will be 98.08\%.
\subsubsection{Alternative Cost Structure within Classifier}
If a major concern was False Positives, we could alter our Recall and Precision values by changing the penalty cost structure within the  CSForest classifier.  In comparison to those values shown in Table \ref{tab:CScost}, by  reducing the penalty cost on False Negatives we may reduce the Recall value, conversely while increasing the False Positive cost  we may increase the Precision value.  As an example of such penalty costs see Table \ref{tab:CScost1a}.

\begin{table}[ht]
\centering
 \caption{ Alternative Penalty Cost Structure for CSForest. }
\begin{tabular}{l|l|c|c|c}
\multicolumn{2}{c}{}&\multicolumn{2}{c}{Actual Value}&\\
\cline{3-4}
\multicolumn{2}{c|}{}&Positive&Negative&\multicolumn{1}{c}{}\\
\cline{2-4}
\multicolumn{1}{c|}{Predicted}& Positive & $TP=1$ & $FP = 2$ & \\[0.2ex]
\cline{2-4}
\multicolumn{1}{c|}{Value} & Negative & $FN = 2$ & $TN = 0$ & \\[0.2ex]
\cline{2-4}

\end{tabular}

 %\caption{ Penalty Cost Structure for CSForest. }
\label{tab:CScost1a}

\end{table}
Applying the CSForest classifier to the  transformed data for the  individuals No. 1 and No. 8, but
using the cost structure as shown in Table \ref{tab:CScost1a},  we note from the results in Table \ref{table:comp2a} that the Recall, Precision and f-score have altered when compared to Table \ref{table:comp}.

\begin{table}[ht]
\caption{Statistical results via Individual  with Alternative Cost Structure}
\label{table:comp2a}
\begin{center}
%\resizebox{\textwidth}{!}{%
\begin{tabular}{ |c|c||c||c|c||c||  }
 \hline
 \multicolumn{6}{|c|}{\bfseries Our Proposed method with Costs Altered} \\
 %\multicolumn{10}{|c|}{\bfseries method numbers \hspace{3.4cm} method numbers}\\
 %\multicolumn{10}{|l|} {\hspace{4.2cm}\textbf{}\hspace{5cm}\textbf{Method Number}} \\
% \multicolumn{5}{l}{\hspace{3.9cm}\textbf{Method Number}}&\multicolumn{5}{l}{\hspace{3.9cm}\textbf{Method Number}}\\
 \hline
 \textit{Ind. No.}& Metric & Result &\textit{Ind. No.}& Metric & Result \\
 \hline
  \textit{1} &Accuracy\%&   98.89 & \textit{8} & Accuracy\%& 96.36 \\
      &Recall& 0.024&     &Recall&  0.191  \\
     &Precision &  1.0&              &Precision &1.0 \\
    & f-score   &  \textbf{0.048} &   & f-score   &\textbf{0.321}  \\
    \hline

 \hline
\end{tabular}
\end{center}
\end{table}

 \subsection{Comparative methods}\label{sub:CopM}  We compared our classification results to two other published methods, using the exact same raw data we had used,  i.e one second epochs from the MIT data set,  
  with results mentioned in Section \ref{sub:results}, Tables \ref{table:fscore}, \ref{table:compA} \& \ref{table:comp}.
  The two works we compared to are as follows:
  \begin{itemize}
  \item [\textbf{Method 1.}] \cite{sidd:2018} used nine statistical parameters, i.e. (Min, Max, \dots, Skewness)  derived from each electrode signal to build the attribute space, resulting in $ 9 \times 23  =  207$ attributes  plus label per record, then applied an ensemble classifier to data, in this case being SysFor \citep{islam:2012}. 
  \item [\textbf{Method 2.}] \cite{Chen:2017}, utilised  the DWT   to decompose the electrode signal into different frequency bands, selected a subset of these frequency bands and  derived statistical parameters from the wavelet coefficients in the chosen frequency bands (wavelet decomposition scales) then applied the classifier SVM, (via a leave one out methodology), to their transformed data.  This was  repeated for different mother wavelets,  (i.e. Daubechies, Haar,  Symlets, etc ...).
  For comparison we used this publication's methodology for the \textit{Haar} wavelet, being six frequency bands together with three statistical parameters: Minimum, Standard deviation and Skewness. This resulted in $6 \times 3 \times 23 = 414$ attributes plus label  per record. 
\end{itemize}

 \subsection{Application of our proposed Method in the The Experiments}
 For this experiment, we used open source software, being R \citep{rcran} for all Graph Theory methods, Wavelet and Statistical analysis.  WEKA \citep{wekas} was utilised for the  Classification step.
 
 \subsubsection{Step 1: Preparation of Attributes  from statistical parameters of reconstructed signal.} \label{sub:ExStep1}
 From the initial  data that had been  segmented into one second epochs, we initially applied the wavelet transform\footnote{Using wavelet filter ``s8" which had been previously used successfully in \cite{brain:2012} \& \cite{Chen:2017}.}. As the original data had been sampled at 256Hz, using the MODWT decomposition we could decompose the signal into  six  frequency bands to match to the advised bandwidths, %EEG waves are conventionally classified into delta, theta, alpha and beta  and gamma waves \citep{Jacob:2018}
 our decomposition/frequency bandwidth range shown in Table \ref{tab:freq}.
 \par
\begin{figure}[ht]
    \centering
    \subfloat[\centering Individual 1 Electrode T7, Energy across Wavelet levels,  Non-Seizure state]{{\includegraphics[width=7cm,height=7.7cm]{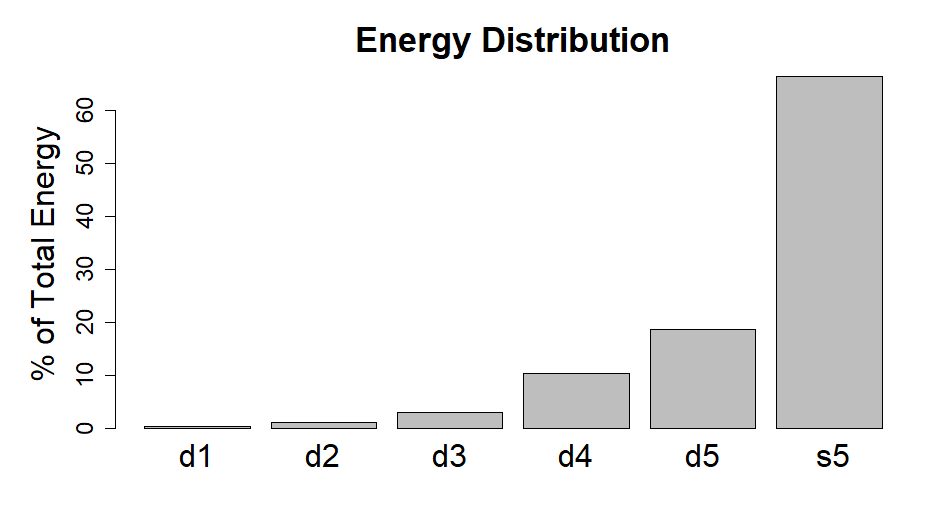} }}%
    \qquad
    \subfloat[\centering Individual 1 Electrode T7, Energy across Wavelet levels,  Seizure state]{{\includegraphics[width=7cm,height=7.7cm]{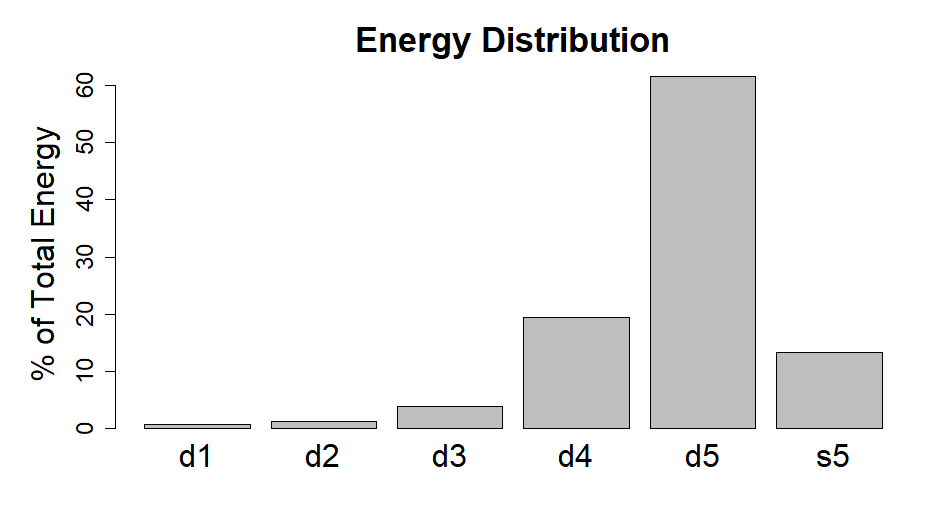} }}%
    \caption{MODWT Decomposition of 1 second Epochs }%
    \label{fig:MODWTD}%
\end{figure}

  We  observed that after such wavelet decomposition, only   minor signal energy  was apparent in the first two wavelet decomposition bands\footnote{These sub-bands are also called Crystals} $\widetilde{\mathcal{D}_1}$ \& $\widetilde{\mathcal{D}_2}$, which represent 32Hz to 128Hz. This was noticed across  many electrodes, both in non-seizure and seizure states, as an example from one individual for one electrode see Figure \ref{fig:MODWTD}.

 %\begin{center}
\begin{table}[ht]
\caption{Frequency Band decomposition}
\label{tab:freq}
\centering
 \begin{tabular}{||c| c c ||} 
 \hline
 Level & Frequency & Band  \\ [0.4ex] 
 \hline\hline
\raisebox{-0.5ex}[10pt]{$\widetilde{\mathcal{D}_1}$} & 64-128Hz & High Gamma  \\ [.5ex]
 \hline
 \raisebox{-0.5ex}[10pt]{$\widetilde{\mathcal{D}_2}$} & 32-64Hz & Gamma \\ [.5ex]
 \hline
 \raisebox{-0.5ex}[10pt]{$\widetilde{\mathcal{D}_3}$} & 16-32Hz & Beta  \\ [.5ex]
 \hline
  \raisebox{-0.5ex}[10pt]{$\widetilde{\mathcal{D}_4}$} & 8-16Hz & Alpha  \\ [.5ex]
 \hline
  \raisebox{-0.5ex}[10pt]{$\widetilde{\mathcal{D}_5}$} & 4-8Hz & Theta \\  [.5ex]
 \hline 
  \raisebox{-0.5ex}[10pt]{$\widetilde{\mathcal{S}_5}$} & 0-4Hz & Delta  \\ [0.5ex]
 \hline
\end{tabular}
\end{table}
Signal energy at each sub-band (\textit{Crystal}), e.g. $\widetilde{\mathcal{D}_4} \sim$  Alpha, is defined similarly as in total energy in signal (\textit{time series}), see  Definition \ref{energy}, Appendix \ref{Defs}

As an example: Energy of Alpha sub-band:
\begin{equation} \label{band:eneg}
Energy|\widetilde{\mathcal{D}_4}| = \sum_{i=1}^n |d_{i,4}|^2 \text{ ,where } d_{1,4}, d_{2,4}, \ldots, d_{n,4}  \in \widetilde{\mathcal{D}_4}
\end{equation}

 After eliminating the the first two wavelet levels (Crystals), we then applied the inverse wavelet transform to reconstruct the signal, where the non-relevant frequency ranges had been omitted. 
 \paragraph{ \textit{feature extraction.}}\label{sub:MODWTfeat}
After application of the inverse wavelet transform on the reduced set of wavelet coefficients, for the epoch, this resulted in  a  time series of length 256 points (1 second epoch). For each of these reconstructed signals, per epoch  we derived the statistical parameters and followed
the procedure as outlined in in Section \ref{sub:step1}.
This provides in total for the 23 electrodes, 115 attributes for each epoch. Table \ref{tab:ExAtt1} is an example of the output derived here..% plus attached label for epoch under consideration. 

\subsubsection{Step 2: Preparation of Attributes from Connectivity of Electrodes} \label{sub:WaveVar}
By applying the MODWT to our data, using same wavelet filter as in Section \ref{sub:ExStep1}, for the  time series from  each electrode, per epoch  we derived wavelet covariance and hence wavelet correlation between each  time series. We then constructed an adjacency matrix  highlighting the direct connections between the electrodes. This was performed using only the wavelet coefficients from the first level of MODWT decomposition, i.e. $d_{i,1} \in \widetilde{\mathcal{D}_1} : i \le n$ and setting a lower bound on the correlation value  of 0.125. Using the lower bound value we determined if  a connection exists between the electrodes if the correlation value is greater than or equal to the lower bound\footnote{The value 0.125 was chosen as with wavelet decomposition level 1, this provided  over 150 electrode connections during the non-seizure state and over 30 connections during the seizure state, per epoch.}.  We represented those connections in  an  adjacency matrix, as described in Section \ref{sub:WaveVar1}.

%\begin{figure}[h]
 %   \centering
  %  \includegraphics[scale=.4]{Net2.png}
  %  \caption{ A Visual  representation of Table \ref{tab:ExAdl}.}
  %  \label{fig:netEXlex}
%\end{figure}

\paragraph{\textit{feature extraction}}\label{sub:WaveVarFeat}  
From the derived adjacency matrix we calculate the sum of the rows.
For each electrode in our network we now have a degree of connections or number of direct connections during the epoch.
From the row sums \textit{(or degree of connection)}  we derive the statistical parameters and follow the procedure  as described in Section \ref{sub:WaveVar1}.  For each epoch this provides us with another five attributes for each record. Table  \ref{tab:ExAtt2a} is an example of this.\par
The data initially  chosen consisted of results from 23 electrodes or channels.  As one channel was the reverse of the other, ie channel 3: T7-P7, channel 19:  P7-T7, we dropped channel 19 when calculating the wavelet covariances and hence from the adjacency matrix construction, we had 22 row sums to derive the statistical parameters from, per epoch.

\vspace{1cm}

\subsubsection {Step 3. Preparation of Attributes from Global Efficiency of Electrodes}\label{sub:Eleceffex}
From Section \ref{sub:WaveVar1} where  wavelet covariance was shown to provide a methodology for the  construction of an adjacency matrix derived from the wavelet correlation between the electrodes, we derive how “efficiently” the electrodes are at communicating across their network. %This originates from Graph theory.  For a complete overview of such theory see: \cite{Andra:1991}. \par
%Our sample network, figure \ref{fig:netEXl}, has 7 nodes/vertices and 9 edges (\textit{Each single edge has an assumed unit length of 1}). For this sample network we may calculate the the global efficiency for each node, see definition \ref{def:globeff}.  We demonstrate this for figure \ref{fig:netEXl},
%Table \ref{tab:ExGlobC} displays the reciprocal of the minimum path length between nodes and the sum of such divided by (N-1)  as per definition \ref{def:globeff}.
This provides us with the global efficiency per node, as described in Section \ref{sub:eleceff}

 We performed this calculation  using the adjacency matrix derived in Step 2 of our experiment, (\textit{Section \ref{sub:WaveVar}}),  for each one second  epoch and  22 electrodes (\textit{nodes}).  
 
 \paragraph{\textit{feature extraction}} \label{sub:ElecEffFeat}
 For each electrode in our network we now have an efficiency value, $Eglob_j : 0 \le  Eglob_j \le 1$ or how well that electrode communicates or transfers information within the  network during the epoch. As we have these values, 22 in total, for each epoch, we derive from them the statistical parameters as described in Section \ref{sub:eleceff} . This provides us with an additional five parameters.   We then place these attributes directly following the attributes derived in Step 1 and Step 2 of our method,  see Table \ref{tab:ExAtt3a} as an example. \par  \textit{These three steps combined together provides  a total 115 + 5 + 5 = 125  attributes.}

 \subsubsection{Step 4. Assigning Class labels and Training Classifiers}
From  the initial three steps in our method, we have constructed a new data set  from the original data, where we have transformed our multivariate EEG data into a univariate list of attributes per Epoch and attached labels as described in Section \ref{subsec:classM}, where Table \ref{tab:ExAttall} is an example of this. \textit{Our constructed  database  of 12 patients which equates to approx 42,816 records, each record of 125 attributes plus label.} We applied the CSForest classifier to our data, with the cost structure as outlined in Section \ref{sub:results},  testing each individual against the  training data set constructed from the other 11 individuals in this study, i.e. leave one out cross validation method. This classification  was undertaken via WEKA. The results of the three different methods per individual are shown in Table \ref{table:comp}.

\section{Conclusion} \label{sec:conc}

In this study, we investigated the use of wavelets and their inherent ability to provide analysis into the time frequency space. We utilised  a wavelet  transform, that is defined for all lengths of sample, able to provide selection of relative frequencies as well as deriving correlation between signals to construct different indicators. Together with some basic methodology from graph theory, we applied these  to an open source data set and compared our results with two other published methods which we  applied to exactly the same data.  Our method compared well, providing overall better results, in regards to f-score and Recall. \par Our methodology of utilising three steps \textit{(consisting of Wavelet applications and Graph theory)} to construct the attributes to which the classifier will be applied, does require some additional computation, but returns a similar number of attributes  when compared to usual methods that construct statistical parameters for attributes derived from raw data. However this method returns considerably less attributes when compared to methods deriving statistical parameters from wavelet coefficients across different decomposition levels.\par

During classification we kept the initial penalty cost structure within the cost sensitive classifier unchanged. This could be altered by determining the optimal cost structure each time the training data set is altered or after inclusion of additional data. The chosen data had 22 electrode positions that could be used to derive node correlation and connectivity. This limited the number of connections available at higher levels of wavelet decomposition without considerably lowering the correlation threshold. These higher levels of wavelet decomposition may have been better suited for seizure discrimination however we did not use owing to the limit on  derived connections. For electrode connectivity calculation we used the first Crystal $\widetilde{\mathcal{D}_1}$ as this decomposition level provided a larger number of connections across the electrodes, especially during seizure, in comparison to the other wavelet levels.
\par
Also we did not weight the distance between the nodes, simply gave a derived connection a distance value or path length of one. Further investigation using data sets with many more nodes together with spatial coordinates  which would enable weighting for distance apart, may prove better at determining electrode connectivity especially in the surrounding locality of seizure, to provide enhanced discrimination between  the different EEG states. \par
Conditional upon acceptance, the associated R code will be made available via Github.

% Acknowledgements should go at the end, before appendices and references

%\acks{To be finalised:   We would like to acknowledge support for this project
%from CSU (PhD stipend)
%and the input and guidance  from Joe blogs, Frank jerk as well as the %Multidisciplinary Research Program of the Department mathematics and computer $science.}

 %{\color{red}{end rubbish}} 

%\newpage

\appendix

%\input{7Appen6}
%\section*{Appendix A.}
%\label{app:theorem}

% Note: in this sample, the section number is hard-coded in. Following
% proper LaTeX conventions, it should properly be coded as a reference:

%In this appendix we prove the following theorem from
%Section~\ref{sec:textree-generalization}:

\section{}

%\label{app:theorem}

% Note: in this sample, the section number is hard-coded in. Following
% proper LaTeX conventions, it should properly be coded as a reference:

%In this appendix we prove the following theorem from
%Section~\ref{sec:textree-generalization}:

\subsection{Glossary}
 \begin{multicols}{2}
    \begin{itemize}
        \item [$A_{i,j}$] Output from classifier, per epoch
        %\item [$At_{ij}$] Combined list of attributes
         \item [$Adj_{ee}$] Adjacency Matrix 
                \item [$C_{i,j}$] List of attributes, Step 2
        \item [$\text{DB}_T$] Database of transformed data
        \item [$d_{i,j}$] $i^{th}$ detail wavelet coefficient
                          at $j^{th}$ level
         \item [$\mathcal{D}_j$] $j^{th}$ level of detail coefficients, DWT
         \item [$\widetilde{\mathcal{D}_j}$] $j^{th}$ level of detail coefficients, MODWT
        \item [DWT] Discrete Wavelet transform
          \item [EEG] Electroencephalogram
         \item [$Eglob_i$] Global Efficiency value, $\text{electrode}_i$ 
         \item [FN] False Negative
         \item [FP] False Positive
         \item [$G_{i,j}$] List of attributes, Step 3
          \item [HPF] High Pass filter
          \item[Hz] Hertz,  One cycle per second
          \item [$I_j$] Individual's data 
         \item [$I_{i,j}$] Individual's data by epoch
       \item [$I_{i,j,e}$] Individual's data by epoch by electrode
       \item [$I_{i,j,e,R}$] Individual's reconstructed data 
             \item [LPF] Low Pass filter
        
          \item [MODWT] Maximum Overlap Discrete Wavelet Transform
           \item [MRA] Multi Resolution Analysis
            \item [$\rho_{\tau \zeta}$] Correlation between the two electrodes ($\tau\, , \, \zeta$) per epoch
        \item[Quartiles] values that divide a  list of numbers into quarters  
        \item [$ r_i$] a  row in Adjacency matrix 
          \item [$R_{i,j}$] Reciprocal of shortest path between two electrodes
        \item [$\text{RW}_{i,j,e}$] List of attributes, Step 1
        \item [$s_{iJ}$]  $i^{th}$ smooth wavelet coefficient at $J^{th}$ level
          \item [$\mathcal{S}_J$] $J^{th}$ level of smooth coefficients, DWT
         \item [$\widetilde{\mathcal{S}_J}$] $J^{th}$ level of smooth coefficients, MODWT
                 \item [$Sp_i$] a set of statistical parameters
                   \item[Skewness] Measure of asymmetry of a probability distribution. 
        \item [SVM] Support Vector Machine, a classifier
      
         \item [$t_h$] derived threshold level 
         \item[TN] True Negative
         \item[TP] True Positive
         \item[$\mu$V] MicroVolt, one millionth of a volt
        \item[$\textbf{W}$] A vector of DWT wavelet coefficients 
        \item[$W_j$]  vector of detail DWT coefficients, $j^{th}$ level
         \item[$V_J$] vector of smooth DWT coefficients, $J^{th}$ level
    \end{itemize}
    \end{multicols}
\newpage

\subsection{Definitions}\label{Defs}
\begin{defn}\label{series1}
Let a sequence $ X_1, X_2,\ldots , X_n $ represent  a time series of $n$ elements, denoted as $\{X_t : t = 1,\ldots, n\},\; X_t  \in \mathbb{R}$.
\end{defn}

 \begin{defn}\label{energy}
The energy within a time series  $X_t$ is the sum of the squared norm of $X_t$, 
\begin{equation*}
   \parallel X_t \parallel^2 \equiv \langle X,X \rangle =  \sum \limits_{t = 1}^{n} X^2_t
 \end{equation*}
\end{defn}

\begin{defn} \label{sam:mean}
let $\overline{X}$ denote the sample mean of $X_t$
\begin{equation*}
\overline{X} \equiv \frac{1}{n} \sum \limits_{t=1}^{n} X_t
\end{equation*}
\end{defn}

\begin{defn} \label{sam:var}
let $\hat{\sigma}^2_X$ denote the sample variance of $X_t$ 
\begin{equation*}
\hat{\sigma}^2_X \equiv \frac{1}{n} \sum \limits_{t=1}^{n} (X_t - \overline{X})^2
\end{equation*}
\end{defn}

\begin{defn} \label{sam:covar}
let  cov\{$X_t , Y_t$\} denote the  covariance of two random variate sequences $X_t \text{and}\;  Y_t$
\begin{equation*}
cov\{X_t , Y_t\} = \sum_{t=1}^n \frac{(X_t -\overline{X})(Y_t - \overline{Y}) }{n}
\end{equation*}
\end{defn}

\begin{defn} \label{def:maf}
\begin{eqnarray*}
\text{macro average Recall (MAR)} & = & \frac{R_1 + R_2 + \dots + R_N}{N}\\
\text{macro average Precision (MAP)} & = & \frac{P_1 + P_2 + \dots + P_N}{N}\\
\text{macro average f-score} & = & \frac{2}{\frac{1}{MAR}+\frac{1}{MAP}} = 2 \times \frac{MAR \times MAP}{ MAR + MAP}\\
\end{eqnarray*}
\hspace{11cm}see \cite{DBLP:journals/corr/Torgo14}
\end{defn}

\begin{defn} \label{def:globeff}
In a Graph G (\textit{or network}) the global efficiency measures how the information is propagating  in the entire network. 
Efficiency = the inverse of the harmonic mean of the minimum path length $L_{jk}$ between a node \textit{j} and all the other nodes \textit{k} in the graph.
\begin{equation*}
\text{Eglob}_j = \frac{1}{N-1} \sum_{k \in G} \frac{1}{L_{jk} }
\end{equation*}
as shown in \cite{brain:2012}
\end{defn}

\subsection{Wavelets}
\subsubsection {DWT}\label{App:DWT}   The discrete wavelet transform (DWT) returns a data vector of the same length as that
of the input. Usually, in this new vector many data points are almost zero. This transform provides an additive decomposition 
of $J$+1 levels.
\begin{equation} \label{DWTaddi}
 X_t \equiv \sum^J_{j=1} \mathcal{D}_j + \mathcal{S}_J  
\end{equation}
 where
 \begin{itemize}
     \item $\mathcal{D}_j$  $j^{th}$ level of wavelet detail coefficients, related to changes at scale $j$
     \item $\mathcal{S}_J$ $J^{th}$ level of wavelet smooth coefficients, related  to averages  at scale $J$
     \item $\{X_t : t = 1,\ldots, n\}$ where $n= 2^J$ : $J \in \mathbb{Z^+},\; X_t  \in \mathbb{R}$,
 \end{itemize}
 The DWT  is a linear transform which decomposes $X_t$, into $J$ levels giving $n$ DWT coefficients;  the wavelet coefficients are obtained  by premultiplying $X$  by $\mathcal{W}$ 
\begin{equation}
 \textbf{W} = \mathcal{W}X
 \end{equation}
\begin{itemize}
    \item $\textbf{W}$ is a vector of DWT coefficients ($j$th component is $W_j$)
    \item $\mathcal{W}$ is $n \times n$ orthonormal transform matrix; i.e.,\newline
    $\mathcal{W}^T \mathcal{W} = I_n$, where $I_n$ is $n \times n$ identity matrix
   \item inverse of $\mathcal{W}$ is its transpose,  $\implies  \mathcal{W} \mathcal{W}^T = I_n$
    \item $\therefore \mathcal{W}^T$\textbf{W} = $\mathcal{W}^T \mathcal{W}X = X$
\end{itemize}
\textbf{W} is partitioned into $J + 1$ subvectors
%\[ \textbf{W} = \left[ \begin{tabular}{c} 
%$\textbf{W}_1$ \\
%$\textbf{W}_2$ \\
%{\scriptsize$\vdots$}  \\
%\textbf{W}_j$ \\
%{\scriptsize$\vdots$}  \\
%$\textbf{W}_J$ \\
%$\textbf{V}_J$ \\
%\end{tabular} \right] \]
\begin{itemize}
    \item $\textbf{W}_j$ has $n/2^j$ elements \footnote{ note: $\sum^J_{j=1} \frac{n}{2^j} = \frac{n}{2} + \frac{n}{4} + \dots + 2 + 1 = 2^J -1 = n-1$}
    \item $\textbf{V}_J$ has one element \footnote{If we decompose $X_t$ to level $I \le J$ then $ \textbf{V}_I$  has $n/2^{I}$ elements.}
\end{itemize}

the synthesis equation for the  DWT is:
\begin{equation} \label{equ:syn}
 X =  \mathcal{W}^T \textbf{W} = \left[ \mathcal{W}^T_1, \mathcal{W}^T_2,\ldots, \mathcal{W}^T_J,  \mathcal{V}^T_J \right] \left[ \begin{tabular}{c} 
$\textbf{W}_1$ \\
$\textbf{W}_2$ \\
{\scriptsize$\vdots$}  \\

$\textbf{W}_J$ \\
$\textbf{V}_J$ \\
\end{tabular} \right]    
\end{equation}

 Equation \ref{equ:syn} leads to additive decomposition which expresses $X$ as the sum of $J+1$ vectors, each of which is associated with a particular scale $\mathcal{T}_j$

\begin{equation} \label{DWTadd}
 = \sum^J_{j=1} \mathcal{W}^T_j\textbf{W}_j + \mathcal{V}^T_J\textbf{V}_J  \equiv \sum^J_{j=1} \mathcal{D}_j + \mathcal{S}_J  
\end{equation}

\begin{itemize}
\item $\mathcal{D}_j \equiv \mathcal{W}^T_j\textbf{W}_j$ is portion of synthesis  due to scale $\mathcal{T}_j$, called the \textit{j}th 'detail'
\item $\mathcal{S}_J  \equiv \, \mathcal{V}^T_J\textbf{V}_j$ is a vector called the 'smooth' of the $J$th order 
\end{itemize}

\subsubsection{MODWT} \label{App:moDWT}

However another variant of the wavelet transform exists called the Maximum Overlap Discrete Wavelet Transform (MODWT), where the number of coefficients returned by the transform at each level of decomposition is the same as that of the input.
\begin{itemize}
    \item The MODWT is not orthonormal and is highly redundant.
    \item The MODWT is defined for $X_t$ on all sample sizes, $n$ does not need to be a power of two
    \item The MODWT to level $J$ has $(J + 1) \times n$ coefficients , the DWT has a total of $n$
coefficients for any chosen $J$.
\item  Similarly to the DWT , the MODWT permits an MRA
    \item  Similarly to the DWT , the MODWT  permits an analysis of variance to be based on MODWT coefficients.
    
\end{itemize}
As with the DWT, from the MODWT we may also obtain a scale based additive decomposition. Where the $\, \mathbf{\widetilde{} } \, $ represents  the coefficients are derived from the MODWT. 
\begin{equation} \label{MODWTaddi}
 X_t = \mathcal{\widetilde{W}}^T_j\widetilde{\textbf{W}}_j + \mathcal{\widetilde{V}}^T_J\widetilde{\textbf{V}}_J \equiv  \sum^J_{j=1} \mathcal{\widetilde{D}}_j + \mathcal{\widetilde{S}}_J  
\end{equation}
Also a scale based energy decomposition, (\textit{can be seen as a basis for ANOVA})
\begin{equation}\label{MODWT:Anova}
 \parallel X_t \parallel^2 = \sum^J_{j=1} \parallel\widetilde{\textbf{W}}_j\parallel^2 + \parallel\widetilde{\textbf{V}}_J\parallel^2
\end{equation}
We may obtain a decomposition of the sample variance, from Definition \ref{sam:var}, it follows:
\begin{equation}
\hat{\sigma}^2_X = \frac{1}{n} \sum_{j=1}^J  \parallel\widetilde{\textbf{W}}_j\parallel^2  + \frac{1}{n}\parallel\widetilde{\textbf{V}}_J\parallel^2 - \overline{X})^2 \equiv \sum_{j=1}^J \nu_X^2(\mathcal{T}_j)
\end{equation}
% Hence wavelet variance performs analysis based upon differences between averages over the differing scales.
We may consider a wavelet variance $\nu_X^2(\mathcal{T}_j)$  as a portion of $\hat{\sigma}^2_X$ related  to changes in averages over scale $\mathcal{T}_j$ i.e., scale by scale analysis of variance. Or the wavelet variance for scale $\mathcal{T}_j$ may be defined as the variance of $\widetilde{W}_j$
\begin{equation}\label{wavVar}
\nu^2_{X}(\mathcal{T}_j) \equiv \text{var}\{\widetilde{W}_j\}
\end{equation}

\subsubsection{Wavelet correlation}\label{sub:wavecorr}
If we have two sequences  (or signals) $X_t \, \& \, Y_t$ as per Definition \ref{series1}, then with wavelet coefficients $\{\widetilde{W}_{X,j} \}$  and $\{\widetilde{W}_{Y,j} \}$, we have: 
%\begin{defn}\label{wave:cov}
wavelet covariance 
\begin{equation} \label{wavCoVar}
\nu_{XY}(\mathcal{T}_j) \equiv cov \{\widetilde{W}_{X,j},\widetilde{W}_{Y,j} \}
\end{equation}
%\end{defn}
which similar to wavelet variance, provides a decomposition at scale $\mathcal{T}_j$,  of the covariance between $\{X_t\}$ and $\{Y_t\}$:
\begin{equation}\label{covarXY}
\sum_{j=1}^\infty \nu_{XY}(\mathcal{T}_j) =  cov \{X_t, Y_t\}
\end{equation} 

Wavelet covariance may be standardised. which yields, wavelet correlation:
\begin{equation}\label{wave:corel}
\rho_{XY} \equiv \frac{cov\{\widetilde{W}_{X,j}, \widetilde{W}_{Y,j}\}}{(var\{\widetilde{W}_{X,j}\}, var\{\widetilde{W}_{Y,j}\})^{1/2}} = \frac{\nu_{XY}(\mathcal{T}_j)}{\nu_X(\mathcal{T}_j)\nu_Y(\mathcal{T}_j)}
\end{equation}

where $\nu_X(\mathcal{T}_j)\; \text{and} \; \nu_Y(\mathcal{T}_j)$ are the wavelet variances for $\{X_t\} \;\text{and} \; \{Y_t\}$,
\cite[see][Chap 8]{percival_walden_2000}.

%\vskip 0.2in
%\bibliography{8sample}

 %\bibliographystyle{splncs04}

\end{document}